\documentclass[aps,superscriptaddress,eqsecnum,showpacs]{revtex4}
\usepackage{graphicx,amssymb,amsbsy,bm,amsmath}
\usepackage{hyperref}

\newcommand{\intp}{\ensuremath{\int\frac{d^3 p}{(2\pi)^3}}}
\newcommand{\intL}{\ensuremath{\int_0^\Lambda}}

\newcommand{\bx}[1][]{\ensuremath{\mathbf{x}{#1}}}
\newcommand{\bk}{\ensuremath{\mathbf{k}}}
\newcommand{\bp}{\ensuremath{\mathbf{p}}}
\newcommand{\bq}{\ensuremath{\mathbf{q}}}

\newcommand{\xt}[1][]{\ensuremath{\mathbf{x}{#1},t{#1}}}
\newcommand{\kt}{\ensuremath{\mathbf{k},t}}

\newcommand{\nn}{\nonumber}

\newcommand{\be}{\begin{equation}}
\newcommand{\ee}{\end{equation}}
\newcommand{\bea}{\begin{eqnarray}}
\newcommand{\eea}{\end{eqnarray}}

\begin{document}
\title{Nonequilibrium pion dynamics near the critical point
in a constituent quark model}
\author{D. Boyanovsky}
\email{boyan@pitt.edu} \affiliation{Department of Physics and
Astronomy, University of Pittsburgh, Pittsburgh, Pennsylvania
15260, USA}\affiliation{LPTHE, Universit\'e Pierre et Marie Curie
(Paris VI) et Denis Diderot (Paris VII), Tour 16, 1er. \'etage, 4,
Place Jussieu, 75252 Paris, Cedex 05, France}
\author{H. J. de Vega}
\email{devega@lpthe.jussieu.fr} \affiliation{LPTHE, Universit\'e
Pierre et Marie Curie (Paris VI) et Denis Diderot (Paris VII),
Tour 16, 1er. \'etage, 4, Place Jussieu, 75252 Paris, Cedex 05,
France}\affiliation{Department of Physics and Astronomy,
University of Pittsburgh, Pittsburgh, Pennsylvania 15260, USA}
\author{S.-Y. Wang}
\email{swang@lanl.gov}
\affiliation{Theoretical Division, MS B285, Los Alamos
National Laboratory, Los Alamos, New Mexico 87545, USA}
\date{\today}

\begin{abstract}
We study static and dynamical critical phenomena of chiral
symmetry breaking in a two-flavor Nambu--Jona-Lasinio constituent
quark model. We obtain the low-energy effective action for scalar
and pseudoscalar degrees of freedom to lowest order in quark loops
and to quadratic order in the meson fluctuations around the mean
field. The \emph{static} limit of critical phenomena is shown to
be described by a Ginzburg-Landau effective action including
\emph{spatial} gradients. Hence \emph{static} critical phenomena
is described by the universality class of the O(4) Heisenberg
ferromagnet. \emph{Dynamical} critical phenomena is studied by
obtaining the equations of motion for pion fluctuations. We find
that for $T<T_c$ the are stable long-wavelength pion excitations
with dispersion relation $\omega_{\pi}(k)=k$ described by isolated
pion poles. The residue of the pion pole vanishes near $T_c$ as $Z
\propto 1/|\ln(1-T/T_c)|$ and long-wavelength fluctuations are
damped out by Landau damping on a time scale
$t_\mathrm{rel}(k)\propto 1/k$, reflecting \emph{critical slowing
down} of  pion fluctuations near the critical point. At the
critical point, the pion propagator features mass shell
logarithmic divergences which we conjecture to be the harbinger of
a (large) dynamical anomalous dimension. We find that while the
\emph{classical spinodal} line coincides with that of the
Ginzburg-Landau theory, the growth rate of long-wavelength
spinodal fluctuations has a richer wavelength dependence as a
consequence of Landau damping. We argue that Landau damping
prevents a \emph{local} low energy effective action in terms of a
derivative expansion in real time at least at the order studied.
\end{abstract}



\maketitle

\section{Introduction}

The spontaneous breaking of chiral symmetry in the QCD vacuum has
fundamental significance in understanding the nonperturbative
nature of hadron dynamics. In QCD with two massless quark flavors
at zero baryon density, the chiral phase transition is expected to
be second order and to be described by the universality class of
the three-dimensional O(4) Heisenberg
model~\cite{pisarski,wilraja}. Recent lattice simulations for
two-flavor QCD in the chiral limit at finite temperature and
vanishing baryon chemical potential seems to indicate that chiral
symmetry is restored above a transition temperature $T_c\sim 170$
MeV, whereas the demonstration of the expected O(4) universality
class still is somewhat ambiguous~\cite{karsch1}.

In ultrarelativistic heavy ion collisions the current theoretical
understanding suggests that a thermalized and almost baryon-free
quark-gluon plasma (QGP) may be formed at a time scale of order 1
fm/c with an initial temperature  larger than the transition
temperature. This quark-gluon plasma then expands hydrodynamically
and cools almost adiabatically, until the transition temperature
is reached at a time scale $10-50$ fm/c depending on the initial
temperature~\cite{qgp}. At BNL Relativistic Heavy Ion Collider
(RHIC) and CERN Large Hadron Collider (LHC) initial temperatures
of the quark-gluon plasma as high as $T\sim 300$ and 500 MeV,
respectively, are expected to be achieved in the most central
collisions, hence providing the experimental possibility to study
the restoration of chiral symmetry and its subsequent spontaneous
breaking as the plasma expands and cools.

While lattice QCD furnishes a nonperturbative tool to study the
thermodynamic equilibrium aspects of the chiral phase transition,
the nonequilibrium aspects which are of fundamental importance and
ubiquitous in the formation and evolution of the QGP in
ultrarelativistic heavy ion collisions cannot be accessed with
this approach. They have to be studied within the framework of
nonequilibrium field theory. However, due to the rich
nonperturbative structure of QCD current theoretical
investigations are limited only to models that incorporate the
relevant symmetries and low-energy effective field theories of
QCD. Some popular models of chiral dynamics are the linear sigma
model~\cite{gellmann}, the Gross-Neveu model~\cite{GN}, and the
Nambu--Jona-Lasinio model~\cite{NJL}. The thermodynamic
equilibrium aspects of these models at finite temperature and/or
quark chemical potential have been studied extensively in the
literature~\cite{klevansky,hatsuda,schwarz,chodos1}, in particular
lattice gauge theory has provided an impressive body of results on
the phase diagram for temperature and chemical potential (see for
example\cite{fodor,philipsen,forcrand,karsch}. Nevertheless, a
microscopic understanding of their dynamical nonequilibrium
aspects has just begun to emerge.

 It has been recently argued that
nonequilibrium dynamical phenomena near the critical point such as
spinodal decomposition, critical slowing down and formation of
correlated pion domains~\cite{wilraja,DCC} may lead to distinct
event-by-event observables in the pion
distribution~\cite{rajagopal} as well as in the low-energy
spectrum of photons~\cite{fotboy}. The nonequilibrium dynamics in
the linear sigma model has been used to study the possible
formation of  large correlated domains of disoriented chiral
condensate (DCC) in ultrarelativistic heavy ion
collisions~\cite{DCC}. Thus non-equilibrium phenomena associated
with chiral symmetry breaking may have important observational
consequences in ultrarelativistic heavy ion collisions.

While several studies of the nonequilibrium dynamics focused on
either a discrete chiral symmetry~\cite{chodos2} or a stochastic
description of long-wavelength fluctuations above the critical
point in the chirally symmetric phase~\cite{koide}, the dynamics
of long-wavelength fluctuations at or below the critical point has
begun to receive attention only recently.

Critical slowing down and the relaxation of long-wavelength
fluctuations at or near the critical temperature has been studied
in the large-$N$ limit of the O($N$) linear sigma model as well as
by implementing a finite-temperature renormalization group in
Ref.~\cite{boyan}. The results of the renormalization group
indicate a new dynamical critical exponent $z\approx 1+1/27$ which
results in a dispersion relation for long-wavelength critical
fluctuations of the form $\omega(k) \propto k^z$ with vanishing
group velocity and critical slowing down in the long-wavelength
limit~\cite{boyan}. This study focused on long-wavelength
fluctuations for $T\geq T_c$ but did not address the relaxation
dynamics of pions below the critical temperature.

In Ref.~\cite{son} the real-time propagation of pions near the QCD
chiral phase transition was studied in an effective chiral theory
which in principle is obtained by integrating out the quark (and
gluon) degrees of freedom of QCD. The effective low-energy theory
proposed in Ref.~\cite{son} on the basis of symmetries, is
\emph{local} in the sense that the terms that describe the
space-time variations of the pion field are written in a gradient
expansion up to second order in space and time derivatives. Based
on this effective chiral action, the conclusion presented in
Ref.~\cite{son} was that the real-time propagation of soft pions
below the critical point can be solely expressed in terms of
\emph{static} correlation functions. An important result of such
analysis is that the pion velocity decreases near the phase
transition. While the results of this study apply to temperatures
below the critical and cannot be directly compared to those of
Ref.~\cite{boyan} (which apply either at or above the critical
temperature), there does seem to be a discrepancy between these
results at least if the results of Ref.~\cite{son} are
extrapolated to the critical point. In particular the study in
Ref.~\cite{boyan} suggests that there is no \emph{local}
description of the dynamics in the low-energy effective theory,
namely the space-time variation of long-wavelength fluctuations
cannot be described in terms of a low-energy effective theory with
a simple derivative expansion for the terms that determine
spatiotemporal variations. In the analysis of Ref.~\cite{boyan}
this breakdown of locality near the critical point emerges in the
form of logarithmic singularities near the mass shell of the
single (quasi)particle excitations, which a resummation via the
renormalization group translates into the anomalous dimension
$z$~\cite{boyan}.

\textbf{Goals of this article}. In Refs.~\cite{son,boyan} the
starting point of the study was a scalar field theory with a
Lagrangian density that describes the spatiotemporal variations in
a local manner in terms of a derivative expansion. In
Ref.~\cite{boyan} the field theory studied was the linear sigma
model, while in Ref.~\cite{son} it was the nonlinear sigma model.
In both cases these are low energy effective field theories where
the terms in the Lagrangian that describe the spatiotemporal
variations are local derivative terms.

In this article we investigate the non equilibrium aspects of the
chiral phase transition \emph{below} the critical temperature in
the chirally broken phase in the two-flavor Nambu-Jona-Lasinio
(NJL) constituent quark model in the strict chiral limit of
massless (up and down) quarks. In this model scalar and
pseudoscalar mesons arise as bound states of quarks and are not
present as elementary fields in the Lagrangian. Their dynamics is
completely determined by the underlying dynamics of quarks. The
effective action for the meson fields is obtained systematically
by integrating out the quark fields, thus the low-energy effective
action for scalar and pseudoscalar mesons can be obtained
explicitly and systematically.

Our main goals are the following:
\begin{itemize}
\item[(i)]{To obtain the low-energy effective action up to
quadratic order in the scalar and pseudoscalar fluctuations around
the mean field in the critical region. Namely, we extract the
Ginzburg-Landau effective theory, both in the static limit as well
as in real time. This is achieved by integrating out the quark
degrees of freedom up to one-loop level. In the NJL model with
$N_f$ flavors and $N_c$ colors, the one loop effective action is
\emph{exact} in the large $N_f N_c$ limit\cite{klevansky,hatsuda}.
Thus although we will focus our discussion on the
phenomenologically relevant case of two flavors and three colors,
the one loop result can be argued to be exact in the limit of
large $N_f N_c$. }

\item[(ii)]{To study the real-time dynamics of long-wavelength
pion fluctuations in the chirally broken phase near and at the
critical temperature. In particular we focus on extracting the
dispersion relation for pions as well as their relaxation.
Furthermore, we also address the dynamics of long-wavelength
spinodal fluctuations and extract their growth rate in the
long-wavelength limit.}

\item[(iii)]{To establish the validity of the assumption on the
universality of the critical phenomena. The
$\mathrm{SU}_V(2)\times \mathrm{SU}_A(2)$ symmetry of the NJL
model suggests that the low-energy effective theory for scalar and
pseudoscalar mesons will be in the same universality class as the
three-dimensional O(4) Heisenberg model~\cite{wilraja}. We study
this aspect both in the static as well as in the real-time
effective action up to the one loop level in quark loops which is
\emph{exact} in the large $N_c N_f$ limit\cite{klevansky}.}
\end{itemize}

The NJL model is clearly not QCD, to begin with it includes the
effect of gluons as an effective four-Fermi interaction and
confinement is not a property of the theory. These shortcomings of
the model notwithstanding, an exhaustive study of its
properties~\cite{hatsuda,klevansky} point to its remarkable
successes in describing the phenomenology of the low-energy
degrees of freedom, namely the pions. In particular important
consistency relations for the two-flavor case such as the
Goldberger-Treiman and Gell-Mann--Oakes--Renner have been shown to
be obeyed by this model (for a detailed analysis
see~\cite{hatsuda,klevansky}). The coupling (and cutoff) of the
model are fixed so that the pion decay constant at zero
temperature is reproduced~\cite{hatsuda,klevansky}. The NJL model
being solely a (constituent) quark model, the meson fields emerge
as quark-antiquark bound states. Hence the \emph{kinetic} terms of
the meson fields in the effective low-energy field theory are
completely determined by quark loops. Thus the static and
real-time aspects of fluctuations beyond the mean field are
completely determined by the correlation functions of quark
 operators.

This article is organized as follows. In Sec.~\ref{sec:model} we
briefly review the two-flavor NJL constituent quark model and to
focus our comparative study. For comparison, we  review as well
the Ginzburg-Landau description of the O(4) linear sigma model. In
Sec.~\ref{sec:stat} we obtain the meson self-energies to lowest
order in the quark-loop expansion and establish the low-energy
effective action in the static limit. In Sec.~\ref{sec:realtime}
we study the dynamics of long-wavelength pion excitations near the
critical point by obtaining and solving the equations of motion
for small amplitude fluctuations around the mean field
configuration. Our conclusions, summary of main results and
further questions are presented in Sec.~\ref{sec:conclu}.

\section{The Nambu--Jona-Lasinio model and the linear sigma model}\label{sec:model}

The constituent quark model that we consider in this article is
the NJL model in the strict chiral limit (vanishing current quark
masses) for $N_c $ colors and $N_f $ flavors. This model is
defined by the Lagrangian density
\begin{equation}
\mathcal{L}_\mathrm{NJL}=\bar{\psi}i\!\!\not\!\partial\psi+
\frac{G^2}{2}[(\bar{\psi}\psi)^2+(\bar{\psi}i\gamma^5
\boldsymbol{\tau}\psi)^2],\label{NJLlag}
\end{equation}
where $\psi$ is column spinor describing Dirac quark fields with
$N_c $ colors and $N_f $ flavors, $G$ is the dimensionful
four-fermion coupling, and $\tau^i$ ($i=1,2,3$) are proportional
to the Pauli matrices and normalized as
\begin{equation}
\mbox{Tr}\,\tau^i \tau^j = N_f \,\delta^{ij},
\end{equation}
with $N_f $ the number of quark flavors. This model has been
studied in detail~\cite{hatsuda,klevansky} and found to describe
fairly well the low-energy phenomenology of pions, which are the
lightest hadronic excitations in QCD. The Lagrangian density
$\mathcal{L}_\mathrm{NJL}$ is invariant under axial-vector
$\mathrm{SU}(2)_V\times\mathrm{SU}(2)_A$ transformations and can
be regarded as the quark part of a low-energy effective theory of
QCD after integrating out the gluon degrees of freedom. The
contact four-fermion interaction is not confining and renders the
NJL model non-renormalizable. A regularization scheme must
therefore be employed to cutoff the divergent quantities that
occur, and the model must be understood as a cutoff effective
field theory.

In the literature, various regularization schemes have been
adopted in the NJL model~\cite{hatsuda,klevansky}. In the
finite-temperature case under study the noncovariant and
physically intuitive three-momentum cutoff scheme is the most
suitable one, where a three-momentum cutoff on the loop momenta is
imposed on all divergent integrals. Consequently, the two-flavor
NJL model in the chiral limit has two parameters: the coupling
constant $G$ and the cutoff $\Lambda$. The ultraviolet cutoff
$\Lambda$ determines the range of validity of the low-energy
effective theory.

We follow Hatsuda et al.~\cite{hatsuda}, who fit the values of the
four-fermi coupling $G$ and of the noncovariant three-momentum
cutoff $\Lambda$ to reproduce the phenomenological parameters
corresponding to the pion decay constant $f_\pi=93$ MeV and the
quark condensate density per flavor $\langle\bar{u}u\rangle
=\langle\bar{d}d\rangle=(-250\;\mathrm{MeV})^3$ in
vacuum~\cite{hatsuda,klevansky}. These authors
found~\cite{endnote}
\begin{equation}
\Lambda = 631\,\mbox{MeV},\quad G^2\Lambda^2=
4.04.\label{parameters}
\end{equation}
We note that at zero temperature the presence of a noncovariant
three-momentum cutoff breaks Lorentz invariance explicitly, hence
care must be taken in implementing the cutoff in order to preserve
the consequences of the chiral symmetry breaking in the (Lorentz
invariant) full theory, viz, QCD. We will address this point in
detail later in our discussion below.

Following the standard Hubbard-Stratonovich procedure by
introducing the composite scalar and pseudoscalar fields
\begin{equation}
\sigma=-G\bar{\psi}\psi,\quad\pi^i=-G\bar{\psi}i\gamma^5\tau^i\psi,
\end{equation}
respectively, we can rewrite $\mathcal{L}_\mathrm{NJL}$ as
\begin{equation}
\mathcal{L}_\mathrm{NJL}=\bar{\psi}i\!\!\not\!\partial\psi-
\frac{1}{2}(\sigma^2+\boldsymbol{\pi}^2)-
G\bar{\psi}(\sigma+i\gamma^5
\boldsymbol{\tau}\cdot\boldsymbol{\pi})\psi,\label{L}
\end{equation}
where $\boldsymbol{\pi}=(\pi^1,\pi^2,\pi^3)$. We note that
$\sigma$ and $\pi$ at tree level are not dynamical fields as there
are no corresponding kinetic terms in the Lagrangian
density~\eqref{L}.

The focus of our study is to compare the static and dynamical
critical phenomena near the critical point obtained from the
effective field theory of scalar and pseudoscalar mesons (pions)
after integrating out the quarks, and the O(4) linear sigma model
which has been argued to describe the universality class for
two-flavor QCD~\cite{pisarski,wilraja}. The Ginzburg-Landau (GL)
description of the linear sigma model (LSM) near the critical
point is based on the Lagrangian
density~\cite{pisarski,wilraja,DCC}
\begin{equation}
\mathcal{L}_\mathrm{LSM} = \frac{1}{2} \partial_{\mu}\sigma
\partial^{\mu}\sigma + \frac{1}{2} \partial_{\mu}\boldsymbol{\pi}\cdot
\partial^{\mu}\boldsymbol{\pi}-
\frac{A}{2}(T^2-T^2_c)(\sigma^2+
\boldsymbol{\pi}^2)-\frac{B}{4}(\sigma^2+
\boldsymbol{\pi}^2)^2,\label{LSM}
\end{equation}
where the coefficients $A,B >0$ are non-universal and vary slowly
near $T_c$. The $T^2$ term emerges from the one-loop tadpole
contribution to the self-energies which is the leading term at
high temperatures and is local~\cite{kapustabook,spino}.

For $T< T_c$ we take the direction of spontaneous symmetry
breaking to coincide with the scalar component $\sigma$ and split
the scalar field into a mean field $\sigma_0$ and fluctuations
$\delta$, while for simplicity we assume that there is no mean
field along the $\pi$ direction. We write $\sigma = \sigma_0
+\delta$. Thus the linear sigma model or Ginzburg-Landau theory in
the approximation of mean field plus \emph{quadratic} fluctuations
around mean field is described by the Lagrangian density
\begin{eqnarray}\label{NLSMF}
\mathcal{L}_\mathrm{LSM} & = & V(\sigma_0)+\frac{d
V(\sigma_0)}{d\sigma_0}\delta+\frac{1}{2}\left[
\partial_{\mu}\delta\partial^{\mu}\delta +
\frac{d^2 V(\sigma_0)}{d\sigma_0^2}\delta^2\right]+ \frac{1}{2}
\left[\partial_{\mu}\boldsymbol{\pi}\cdot
\partial^{\mu}\boldsymbol{\pi}+\frac{1}{\sigma_0} \frac{d
V(\sigma_0)}{d\sigma_0}\boldsymbol{\pi}^2\right],\nn\\
V(\sigma_0)& = &
\frac{A}{2}(T^2-T^2_c)\sigma^2_0+\frac{B}{4}\sigma^4_0.
\end{eqnarray}
When the order parameter $\sigma_0$ is at its thermodynamic
equilibrium value corresponding to $d V(\sigma_0)/d\sigma_0=0$ the
$\vec{\pi}$ field is massless and describes a triplet of Goldstone
bosons.

The goal of our study is to compare the low-energy effective field
theory of meson fields obtained by integrating out the fermion
fields in the constituent quark model, up to \emph{quadratic}
order in the fluctuations around the homogeneous mean field. This
comparison will allow us to establish the validity of the
universality hypothesis for both static as well as dynamic
critical phenomena at least to lowest order in the fluctuations.

\section{Static Ginzburg-Landau effective theory}\label{sec:stat}

We begin our study by obtaining the GL effective action up to
Gaussian (quadratic) fluctuations around a mean field in the
Matsubara formulation. The continuation to imaginary time that
describes the theory at finite temperature is obtained by taking
$it \rightarrow \tau$ with $0\leq \tau \leq \beta=1/T$ but keeping
the same Dirac gamma matrices as in Minkowski space-time.
Furthermore, we shift the scalar condensate $\sigma$ by a
space-time constant mean field as
\begin{equation}\label{MF}
\sigma(\bx,\tau)= \sigma_0 + \delta(\bx,\tau),
\end{equation}
and define the effective fermion mass as
\begin{equation}\label{mass}
m=G \sigma_0.
\end{equation}
We have chosen the direction of symmetry breaking to coincide with
the $\sigma$ for simplicity, of course we could have chosen mean
fields for the pions $\boldsymbol{\pi}_0$ as well. In such a case,
the underlying $\mathrm{SU}(2)_\mathrm{V}$ isospin symmetry (for
massless quarks) entails that the mass term would be $m= G
\sqrt{\sigma^2_0+\boldsymbol{\pi}^2_0}$. Choosing the $\sigma$
direction does not represent a loss of generality, by invoking the
underlying isospin we can recover the general result  by simply
replacing $\sigma^2_0 \rightarrow
\sigma^2_0+\boldsymbol{\pi}^2_0$.

The fermionic path integral can be carried out with the result for
the total path integral
\begin{eqnarray}
Z &=& \int \mathcal{D}\delta \mathcal{D}\boldsymbol{\pi}
\exp\bigg\{-\beta V \frac{\sigma^2_0}{2} -\int d^3x \int_0^\beta
d\tau \bigg[\sigma_0 \delta(\bx,\tau)+\frac{1}{2}
\delta^2(\bx,\tau)+\frac{1}{2}\boldsymbol{\pi}^2(\bx,\tau)\nn\\
&&-\mathrm{Tr}\ln(i\!\!\not\!\partial_E-m -G \delta-iG\gamma^5
\boldsymbol{\tau}\cdot
\boldsymbol{\pi})\bigg]\bigg\},\label{PFtot}
\end{eqnarray}
where $i\!\!\not\!\partial_E = -\gamma^0 \partial/
\partial \tau + i \boldsymbol{\gamma}\cdot \boldsymbol{\nabla}$
and $V$ is the spatial volume. The fermionic path integral
(partition function)
\begin{eqnarray}
Z[\delta,\boldsymbol{\pi}] & = &
\exp[\mathrm{Tr}\ln(i\!\!\not\!\partial_E-m -G\delta-
iG\gamma^5 \boldsymbol{\tau}\cdot\boldsymbol{\pi})]\nonumber \\
& = &  \int\mathcal{D}\bar{\psi} \mathcal{D}\psi \exp\bigg\{\int
d^3 x \int_0^\beta d\tau
\bar{\psi}(\bx,\tau)\left[i\!\!\not\!\partial_E -m -G
\delta(\bx,\tau)-iG\gamma^5 \boldsymbol{\tau} \cdot
\boldsymbol{\pi}(\bx,\tau) \right]\psi(\bx,\tau)\bigg\}\label{FPF}
\end{eqnarray}
can be written in a more familiar manner as
\begin{equation}\label{expval}
Z[\delta,\boldsymbol{\pi}] = Z_\mathrm{MF} \bigg\langle
T_\tau\exp\bigg\{-G \int d^3x \int_0^\beta d\tau \bar{\psi}(
\delta + i\gamma^5 \boldsymbol{\tau}\cdot \boldsymbol{\pi}) \psi
\bigg\}\bigg\rangle_0,
\end{equation}
where $T_\tau$ is the $\tau$-ordering symbol and $Z_\mathrm{MF}$
is the free-field partition function for $N_c \times N_f $ Dirac
fermions of mass $m=G\sigma_0$, namely
\begin{equation}\label{FinTZfer}
Z_\mathrm{MF} = \mathrm{Tr} e^{-\beta H_0} = \exp\left\{2\beta V
N_c  N_f   \int \frac{d^3 k}{(2\pi)^3}\left[\frac{2}{\beta}
\ln(1+e^{-\beta \omega_k})+\omega_k\right] \right\},
\end{equation}
with $\omega_k = \sqrt{k^2+m^2}$ and $\langle \cdots \rangle_0$
the thermal expectation value in a free field theory of Dirac
fermions with mass $m$, $N_c $ colors and $N_f $ flavors. The
$\mathrm{Tr}\ln[\cdots]$ in eq.~(\ref{PFtot}) can be expanded
systematically in powers of $G$, the effective action in the mean
field approximation plus quadratic fluctuations around the mean
field is obtained by expanding up to $\mathcal{O}(G^2)$.
Equivalently, the expectation value of the exponential in
eq.~(\ref{expval}) is expanded up to $\mathcal{O}(G^2)$ and the
result re-exponentiated. Following these steps we are led to the
finite-temperature effective action up to quadratic order in the
fluctuations of the meson fields $\delta$ and $\pi$. Taking the
spatial Fourier transform as well as the discrete Fourier
transform in terms of Matsubara frequencies (in the interval $0
\leq \tau \leq \beta$)
\begin{equation}
\delta(\bx,\tau)= \frac{1}{\sqrt{V}}\sum_{\nu_n}\sum_{\bk}
\delta(\bk,\nu_n)\, e^{-i\nu_n\tau+i\bk \cdot \bx},\quad\nu_n=2\pi
n/\beta,
\end{equation}
and similarly for $\pi^i$, we find
\begin{eqnarray}\label{Seff}
S_\mathrm{eff}[\delta,\pi;\sigma_0] & = &  \beta V\bigg\{
\mathcal{V}_\mathrm{eff}(\sigma_0)+
\frac{d\mathcal{V}_\mathrm{eff}(\sigma_0)}{d\sigma_0}\delta
(\mathbf{0},0) + \sum_{\nu_n} \int
\frac{d^3k}{(2\pi)^3}\bigg[\frac{1}{2} \delta(\bk,\nu_n)[1+
\Sigma_{\delta}(\bk,\nu_n)]\delta(-\bk,-\nu_n)\nn\\
&& +\frac{1}{2} \pi^i(\bk,\nu_n)[\delta^{ij}+
\Sigma^{ij}_{\pi}(\bk,\nu_n)]\pi^j(-\bk,-\nu_n)\bigg]\bigg\},
\end{eqnarray}
with $\nu_n=2\pi n/\beta$ the bosonic Matsubara frequencies.

The finite-temperature effective potential is given by
\begin{equation}\label{Veff}
\mathcal{V}_\mathrm{eff}(\sigma_0)= \frac{1}{2}\sigma^2_0 - 2 N_c
N_f \int \frac{d^3 k}{(2\pi)^3}\Big[\frac{2}{\beta}
\ln(1+e^{-\beta \omega_k})+\omega_k\Big],
\end{equation}
and its derivative with respect to $\sigma_0$ reads
\begin{equation}\label{derVeff}
\frac{d\mathcal{V}_\mathrm{eff}(\sigma_0)}{d\sigma_0}  = \sigma_0+
G \langle \bar{\psi}\psi \rangle_0 = \sigma_0\left[1-2\,N_c N_f
G^2 \int \frac{d^3 k}{(2\pi)^3}\frac{1-2\,n(\omega_k)}{\omega_k}
\right],
\end{equation}
where $n(\omega_k)= 1/(e^{\beta\omega_k}+1)$. The equilibrium
state of the system is determined by the condition
$d\mathcal{V}_\mathrm{eff}(\sigma_0)/d\sigma_0 = 0$, which leads
to the gap equation
\begin{equation}
\sigma_0+ G \langle \bar{\psi}\psi \rangle_0 =
\sigma_0\left[1-2N_c  N_f G^2  \int \frac{d^3
k}{(2\pi)^3}\frac{1-2n(\omega_k)}{\omega_k} \right]=0.
\end{equation}
For temperatures above the critical $T_c$ (see below) the only
available solution is $\sigma_0=0$ whereas for $T< T_c$ there is a
non-trivial solution which signals spontaneous chiral symmetry
breaking~\cite{hatsuda,klevansky}. We have chosen the direction of
symmetry breaking along the $\sigma$ component of the multiplet
for convenience, since in this case the $\vec\pi$ field describes
the multiplet of Goldstone bosons, namely pions.

The meson self-energies are given by
\begin{eqnarray}
\Sigma_{\delta}(\bk,\nu_n) & = & - G^2\int d^3 x \int_0^\beta
d\tau e^{i\nu_n\tau} \langle T_\tau \rho(\bx,\tau) \rho(\mathbf{
0},0)\rangle^c_0, \nn\\
\Sigma^{ij}_{\pi}(\bk,\nu_n) & = & - G^2\int d^3 x \int_0^\beta
d\tau e^{i\nu_n\tau} \langle T_\tau \rho^i_5(\bx,\tau)
\rho^j_5(\mathbf{0},0)\rangle^c_0, \label{SEpi}
\end{eqnarray}
where the superscript $c$ and subscript $0$ refer to the connected
correlation function in the free field theory and the scalar and
chiral densities are given by
\begin{eqnarray}
\rho(\bx,\tau) & =  &  \bar{\psi}(\bx,\tau)\psi(\bx,\tau),
\nn\\
\rho^i_5(\bx,\tau) & =  &  \bar{\psi}(\bx,\tau)i\gamma^5
\tau^i\psi(\bx,\tau).\label{chiral}
\end{eqnarray}
We discuss in detail both the effective potential and the
self-energies separately below.

\subsection{The effective potential and its Ginzburg-Landau
expansion near the critical point}

The gap equation which describes the  thermodynamic equilibrium
state of the system is given by
\begin{equation}
\frac{d\mathcal{V}_\mathrm{eff}(\sigma_0)}{d\sigma_0} = \sigma_0+
G \langle \bar{\psi}\psi \rangle_0 = \sigma_0\left[1-2N_c N_f G^2
\int \frac{d^3 k}{(2\pi)^3}\frac{1-2n(\omega_k)}{\omega_k}
\right]=0.\label{gapeqn}
\end{equation}
There is an obvious solution $\sigma_0=0$ corresponding to
unbroken chiral symmetry, but below the critical temperature there
is another solution with $\sigma_0 \neq 0$ which describes
spontaneous chiral symmetry breaking~\cite{hatsuda,klevansky} and
corresponds to the vanishing of terms in the brackets in
eq.~\eqref{gapeqn}. The critical temperature corresponds to the
vanishing of this nontrivial solution of the gap equation and is
determined by the equation
\begin{equation}
1= \frac{N_c  N_f G^2}{\pi^2} \int^{\Lambda}_0
dk\,k\,\tanh\frac{k}{2T_c} \simeq\frac{N_c N_f
G^2}{2\pi^2}\left[\Lambda^2 - \frac{\pi^2}{3}
T^2_c\right],\label{Tc}
\end{equation}
where we have neglected terms that are exponentially small in the
ratio $\Lambda/T_c$. The critical temperature is thus found to be
given by~\cite{klevansky}
\begin{equation}\label{Tcfin}
T_c \simeq \frac{\sqrt{3}}{\pi} \left[1- \frac{2\pi^2}{N_c N_f
G^2 \Lambda^2 } \right]^{\frac{1}{2}}\Lambda,
\end{equation}
which for $N_f =2$, $N_c =3$ and the values of the parameters
given by (\ref{parameters}) yields $T_c \simeq
150~\mbox{MeV}$~\cite{hatsuda,klevansky}.

For $T \lesssim T_c$ we expand the derivative of the effective
potential (\ref{derVeff}) as a power series expansion in
$\sigma_0$, this is achieved by the following steps: (i) In the
integral in (\ref{derVeff}) write the integrand as
\begin{equation}\label{trick}
\frac{1-2n(\omega_k)}{\omega_k}= \frac{1-2n(k)}{k}+ \left[
\frac{1-2n(\omega_k)}{\omega_k}- \frac{1-2n(k)}{k}\right],
\end{equation}
and the integration of the first term in (\ref{trick}) gives
\begin{equation}
\int^{\Lambda}_0 dk\,k\, \tanh\frac{k}{2T} \simeq
\frac{1}{2}\left[\Lambda^2 - \frac{\pi^2 }{3}T^2\right].
\end{equation} (ii) Use the identity
\begin{equation}
\frac{1-2n(\omega)}{\omega}= 2 T \sum_{\omega_n}
\frac{1}{\omega^2_n+\omega^2},\quad \omega_n = (2n+1)\pi T,
\end{equation}
and expand the difference in the bracket in (\ref{trick}) in a
power series expansion in $m^2$. The result can be integrated term
by term.

Upon using the result (\ref{Tcfin}) for the critical temperature,
we find
\begin{equation}\label{derLG}
\frac{d\mathcal{V}_\mathrm{eff}(m)}{dm}= \frac{N_c N_f}{6}m
\Bigg[T^2-T^2_c + m^2\, F\left(\frac{\Lambda}{2T_c}\right) +
\mathcal{O}(m^4)\Bigg],
\end{equation}
where
\begin{equation}
F(y) = \frac{3}{\pi^2}\int^y_0  \frac{dx}{x}\tanh x .
\end{equation}
 For the relevant case
$N_c =3$ and $N_f =2$ with the values of the parameters given in
(\ref{parameters}) and that for $T_c$ given by eq.~(\ref{Tcfin})
we find $F(\Lambda/2T_c)\approx 0.477$. Integrating this
expression we find the effective potential in the GL region (i.e.,
$|T-T_c|/T_c\ll 1$ and $m/T\ll 1$) to be given by
\begin{equation}
\mathcal{V}_\mathrm{eff}(m) = \frac{N_c N_f }{6}\left[
\frac{m^2}{2} (T^2-T^2_c)+  \frac{m^4}{4}
F\left(\frac{\Lambda}{2T_c}\right)\right]+\mathcal{O}(m^6),\label{VeffLG}
\end{equation}
where the higher order terms are subdominant in the GL region near
the critical temperature. In order to extract the momentum
dependence of the fluctuation contributions to the GL effective
action at finite temperature, we must study the self-energies.

\subsection{Self-energies and dispersion relations}

Anticipating our study of the real-time dynamics, in this section
we summarize important aspects of the self-energies which will be
necessary for the low-energy expansion both in the static as well
as in the dynamical study. Taking the spatial Fourier transform,
the (Matsubara) self-energies in eq.~(\ref{SEpi}) are given by
\begin{equation}\label{correl}
\Sigma_{\delta}(\bk,\nu_n) = -  \int_0^\beta d\tau e^{i\nu_n\tau}
\langle \rho_{\bf k}(\tau) \rho_{-\bf k}(0)\rangle^c_0,
\end{equation}
and similarly for $\Sigma^{ij}_{\pi}$. In what follows when no
confusion occurs, we will not write explicitly the isospin indices
$i,j$ for $\Sigma_{\pi}$ to avoid cluttering of notation. The
correlator in eqn. (\ref{correl}) can be written as a spectral
representation
\begin{eqnarray}
\langle \rho_{\bf k}(\tau) \rho_{-\bf k}(0)\rangle^c_0  & = &
\frac{1}{Z_\mathrm{MF}}\sum_{m,n} e^{-\beta E_n}\langle n|e^{\tau
H_0}\rho_{\bf k}(0)e^{-\tau H_0}|m\rangle\langle m|\rho_{-\bf
k}(0)|n\rangle \nonumber \\
& = & \frac{1}{Z_\mathrm{MF}}\sum_{m,n} e^{-\beta
E_n}e^{(E_n-E_m)\tau}\langle n|\rho_{\bf k}(0)|m\rangle \langle
m|\rho_{-\bf k}(0)|n\rangle
\end{eqnarray}
and a similar representation for the correlator that enters in the
definition of $\Sigma_\pi$ in eq.~(\ref{SEpi}).

The integral over the imaginary time in eq.~(\ref{SEpi}) can now
be done straightforwardly and we find the spectral representation
for the self-energies
\begin{eqnarray}
\Sigma_{\delta}(k,\nu_n) & = & -\frac{1}{\pi}
\int_{-\infty}^{\infty} dk_0
\frac{\mathcal{S}_{\delta}(k,k_0)}{k_0-i\nu_n}, \nn\\
\Sigma^{ij}_{\pi}(k,\nu_n) & = &  -\frac{1}{\pi}
\int_{-\infty}^{\infty} dk_0
\frac{\mathcal{S}^{ij}_{\pi}(k,k_0)}{k_0-i\nu_n}, \label{SEpiSR}
\end{eqnarray}
where
\begin{eqnarray}
\mathcal{S}_{\delta}(k,k_0) & = &   \frac{G^2
\pi}{Z_\mathrm{MF}}\sum_{m,n} e^{-\beta E_n}\langle n|\rho_{\bf
k}(0)|m\rangle \langle m|\rho_{-\bf k}(0)|n\rangle
\delta(k_0-E_n+E_m) (e^{\beta k_0}-1),\nn\\
\mathcal{S}_{\pi}^{ij}(k,k_0) & = &   \frac{G^2
\pi}{Z_\mathrm{MF}}\sum_{m,n} e^{-\beta E_n}\langle
n|\rho^i_{5,\bf k}(0)|m\rangle \langle m|\rho^j_{5,-\bf
k}(0)|n\rangle \delta(k_0-E_n+E_m) (e^{\beta k_0}-1).
\label{delsig}
\end{eqnarray}

With the purpose of studying the real-time dynamics of pion
fluctuations in linear response later and to establish contact
between the real-time dynamics and the Matsubara formulation, let
us introduce the following \emph{real-time} correlation functions
\begin{eqnarray}
\Sigma^>_{\delta}(k,t-t') & = & G^2 \langle \rho_{\bf
k}(t)\rho_{-\bf k}(t')\rangle_0^c,\nn\\
\Sigma^<_{\delta}(k,t-t') & = & G^2 \langle \rho_{-\bf
k}(t')\rho_{\bf k}(t)\rangle_0^c, \label{siggreat}
\end{eqnarray}
where $\rho_{\bf k}(t) = e^{iH_0 t} \rho_{\bf k}(0) e^{-iH_0 t}$,
and similarly for $\Sigma^\lessgtr_\pi$ in terms of the chiral
densities. Introducing a complete set of eigenstates of $H_0$, we
find
\begin{equation}
\Sigma^\lessgtr_{\delta,\pi}(k,t-t') =
\int_{-\infty}^{\infty}d\omega\,\sigma^\lessgtr_{\delta,\pi}(k,\omega)\,
e^{i\omega(t-t')},\label{Sigreps}
\end{equation}
in terms of the spectral densities
\begin{eqnarray}
\sigma^>_{\delta}(k,\omega) & = & \frac{G^2
}{Z_\mathrm{MF}}\sum_{m,n} e^{-\beta E_n}\langle n|\rho_{\bf
k}(0)|m\rangle \langle m|\rho_{-\bf k}(0)|n\rangle
\delta(\omega-E_n+E_m),\nn\\
\sigma^<_{\delta}(k,\omega) & = & \frac{G^2
}{Z_\mathrm{MF}}\sum_{m,n} e^{-\beta E_n}\langle n|\rho_{-\bf
k}(0)|m\rangle \langle m|\rho_{\bf k}(0)|n\rangle
\delta(\omega-E_m+E_n).\label{siggr}
\end{eqnarray}
Upon relabelling the sum indices $m \leftrightarrow n$ in the
expression for $\sigma^<$ above, we find the Kubo-Martin-Schwinger
(KMS) condition~\cite{kapustabook,lebellac}
\begin{equation}
\sigma^<_{\delta}(k,\omega)= \sigma^>_{\delta}(k,-\omega)=e^{\beta
\omega} \sigma^>_{\delta}(k,\omega). \label{KMS}
\end{equation}
Similar expressions are found for
$\sigma^\lessgtr_{\pi}(k,\omega)$ with the scalar densities
replaced by the chiral densities and the appropriate isospin
indices $i,j$.

The retarded self-energy is given by (see below)
\begin{equation}\label{retSE}
\Sigma^\mathrm{ret}_{\delta,\pi}(k,t-t')=
-i\left[\Sigma^>_{\delta,\pi}(k,t-t')-\Sigma^<_{\delta,\pi}(k,t-t')
\right]\Theta(t-t') \equiv
\Sigma_{\delta,\pi}(k,t-t')\Theta(t-t'),
\end{equation}
where the second equivalence \emph{defines}
$\Sigma_{\delta,\pi}(k,t-t')$. Using the spectral representation
of the $\Theta(t-t')$ we find
 \begin{equation}\label{sigi}
\Sigma_{\delta,\pi}(k,t-t') = \int^\infty_{-\infty}
\frac{dk_0}{2\pi} e^{ik_0(t-t')}
\widetilde{\Sigma}^\mathrm{ret}_{\delta,\pi}(k,k_0),
\end{equation}
with
\begin{equation}\label{disret}
\widetilde{\Sigma}^\mathrm{ret}_{\delta,\pi}(k,k_0) =
\int^\infty_{-\infty} d\omega \frac{
\sigma^>_{\delta,\pi}(k,\omega)-
\sigma^<_{\delta,\pi}(k,\omega)}{\omega-k_0+i\epsilon}
\equiv -\frac{1}{\pi}\int^\infty_{-\infty}d\omega
\frac{\mathrm{Im}
\widetilde{\Sigma}^\mathrm{ret}_{\delta,\pi}(k,\omega)}{\omega-k_0+i\epsilon}.
\end{equation}
Using the KMS condition (\ref{KMS}) we find the imaginary part of
the retarded self-energy to be given by
\begin{equation}\label{imSE}
\mathrm{Im}\widetilde{\Sigma}_{\delta,\pi}(k,\omega) = -\pi
[\sigma^>_{\delta,\pi}(k,\omega)-\sigma^<_{\delta,\pi}(k,\omega)]=
\pi (e^{\beta \omega}-1) \sigma^>_{\delta,\pi}(k,\omega).
\end{equation}
Similarly for $\Sigma_{\delta,\pi}(k,t-t')$ introduced on the
right hand side in eq.~(\ref{retSE}) we find
\begin{equation}\label{intrep}
\Sigma_{\delta,\pi}(k,t-t') = -i \int^{\infty}_{-\infty} d\omega
e^{i\omega(t-t')}[\sigma^>_{\delta,\pi}(k,\omega)
-\sigma^<_{\delta,\pi}(k,\omega)] =
\frac{i}{\pi}\int^{\infty}_{-\infty} d\omega
e^{i\omega(t-t')}\mathrm{Im}\widetilde{\Sigma}^\mathrm{ret}_{\delta,\pi}(k,\omega).
\end{equation}
Comparing eq.~(\ref{imSE}) with $\sigma^>_{\delta,\pi}(k,\omega)$
given by eq.~(\ref{siggr}) with $\mathcal{S}_{\delta}(k,k_0)$
given by eq.~(\ref{delsig}) and which enters in the dispersive
representation of the self-energy in eq.~(\ref{SEpiSR}), we see
that
\begin{eqnarray}
\mathcal{S}_{\delta}(k,k_0) & = &
\mathrm{Im}\widetilde{\Sigma}_{\delta}(k,k_0),\nn\\
\mathcal{S}^{ij}_{\pi}(k,k_0) & = &
\mathrm{Im}\widetilde{\Sigma}^{ij}_{\pi}(k,k_0),\label{matsuIMpi}
\end{eqnarray}

Anticipating a study of relaxation in real time as an initial
value problem, we introduce the Laplace transform of
$\Sigma_{\delta,\pi}(k,t)$ introduced on the right hand side in
eq.~(\ref{retSE})
\begin{equation}
\widetilde{\Sigma}_{\delta,\pi}(k,s) = \int^\infty_0 e^{-st}
\Sigma_{\delta,\pi}(k,t)dt = -\frac{1}{\pi}\int^\infty_{-\infty}
\frac{\mathrm{Im}
\widetilde{\Sigma}^\mathrm{ret}_{\delta,\pi}(k,\omega)}{\omega+is}.
\label{LaplaSig}
\end{equation}
Upon comparing this expression for the Laplace transform with the
dispersive representation of the retarded self-energy
(\ref{disret}) as well as the Matsubara correlators
(\ref{SEpiSR}), we find the following relations between the
Fourier transform of the retarded self-energy, the Matsubara
self-energy and the Laplace transform introduced above
\begin{eqnarray}
\widetilde{\Sigma}^\mathrm{ret}_{\delta,\pi}(k,k_0) & = &
\widetilde{\Sigma}_{\delta,\pi}(k,s=ik_0+\epsilon),\nonumber\\
\Sigma_{\delta,\pi}(k,\nu_n) & = &
\widetilde{\Sigma}_{\delta,\pi}(k,s=-\nu_n).\label{anacont}
\end{eqnarray}
These relations will allow us to study  the static aspects in the
Matsubara representation and the real-time dynamics in an unified
manner.

The main point of this analysis and the relationship between the
self-energies, is that we can obtain both the Matsubara as well as
the real-time self-energies directly from the computation of the
self-energy in the Matsubara representation.

\subsection{Calculation of the self-energies and the low-energy expansion}

Having established the relationships between the self-energies in
imaginary time and in real time, the next step is to obtain the
explicit expression of the self-energy. It is more convenient to
obtain the Matsubara self-energy, once it is written as a
dispersion relation we can use the relations given by
eq.~(\ref{anacont}) and obtain the corresponding expression in
real time.

The most straightforward approach to obtaining the Matsubara
self-energy directly in a dispersive form consists of writing the
fermion propagator in the Matsubara representation as a dispersion
relation (with zero chemical potential)
\begin{equation}\label{ferprop}
S(\bp,\omega_m) = \int_{-\infty}^{\infty} dp_0
\frac{\rho(\bp,p_0)}{p_0-i\omega_m},\quad\omega_m = (2m+1)\pi T,
\end{equation}
with the spectral function for free Dirac fermions of mass $m$
given by
\begin{equation}
\rho(\bp,p_0)=
\frac{1}{2}\left[\gamma^0-\frac{\boldsymbol{\gamma}\cdot \bp -
m}{\omega_p}\right]\delta(p_0-\omega_p)+\frac{1}{2}
\left[\gamma^0+\frac{\boldsymbol{\gamma}\cdot \bp
-m}{\omega_p}\right]\delta(p_0+\omega_p).\label{propDR}
\end{equation}
The one-loop self-energies are given by
\begin{eqnarray}
\Sigma_{\delta}(\bk,\nu_n) & = & G^2 T\sum_{\omega_m}
\int\frac{d^3p}{(2\pi)^3} \int_{-\infty}^{\infty} dp_0
\int_{-\infty}^{\infty} dq_0\,\mathrm{tr}\,
\frac{\rho(\bp,p_0)}{p_0-i\omega_m}
\frac{\rho(\bq,q_0)}{q_0-i\omega_m-i\nu_n},\nn\\
\Sigma^{ij}_{\pi}(\bk,\nu_n) & = & G^2 T\sum_{\omega_m}
\int\frac{d^3p}{(2\pi)^3} \int_{-\infty}^{\infty} dp_0
\int_{-\infty}^{\infty} dq_0\,\mathrm{tr}\,i\gamma^5 \tau^i
\frac{\rho(\bp,p_0)}{p_0-i\omega_m}\,i\gamma^5
\tau^j\,\frac{\rho(\bq,q_0)}{q_0-i\omega_m-i\nu_n},\label{sigipi}
\end{eqnarray}
where $\bq=\bp+\bk$. The sum over the fermionic Matsubara
frequencies can be done
straightforwardly~\cite{kapustabook,lebellac} leading to the
dispersive form
\begin{eqnarray}
\Sigma_{\delta}(\bk,\nu_n) & = & G^2 \int\frac{d^3p}{(2\pi)^3}
\int_{-\infty}^{\infty} dp_0 \int_{-\infty}^{\infty} dq_0
\int_{-\infty}^{\infty} dk_0\,\mathrm{tr}
\left[\rho(\bp,p_0)\rho(\bq,q_0)\right]\nn\\
&&\times\frac{n(q_0)-n(p_0)}{k_0-i\nu_n}\,\delta(k_0-q_0+p_0) \nonumber \\
& \equiv  & -\frac{1}{\pi} \int_{-\infty}^{\infty} dk_0
\frac{\mathrm{Im}
\widetilde{\Sigma}^\mathrm{ret}_{\delta}(k,k_0)}{k_0-i\nu_n},\nn\\
\Sigma_{\pi}^{ij}(\bk,\nu_n) & = & G^2 \int\frac{d^3p}{(2\pi)^3}
\int_{-\infty}^{\infty} dp_0 \int_{-\infty}^{\infty} dq_0
\int_{-\infty}^{\infty} dk_0\,\mathrm{tr} \left[i\gamma^5 \tau^i
\rho(\bp,p_0) i\gamma^5
\tau^j \rho(\bq,q_0)\right]\nn\\
&&\times\frac{n(q_0)-n(p_0)}{k_0-i\nu_n}\,\delta(k_0-q_0+p_0)\nonumber \\
& \equiv  & -\frac{1}{\pi} \int_{-\infty}^{\infty} dk_0
\frac{\mathrm{Im}
\widetilde{\Sigma}^{\mathrm{ret},ij}_{\pi}(k,k_0)}{k_0-i\nu_n},
\label{sigipi2}
\end{eqnarray}
where on the right hand sides of the equations above we have used
the relations given by eqns. (\ref{LaplaSig},\ref{anacont})
established above .

Using the identities
\begin{eqnarray}
\omega_q\omega_p\pm\bp\cdot\bq\mp m^2&=&\pm
\frac{1}{2}[(\omega_q\pm\omega_p)^2-k^2-4m^2],\nn\\
\omega_q\omega_p\pm\bp\cdot\bq\pm m^2&=&\pm
\frac{1}{2}[(\omega_q\pm\omega_p)^2-k^2],
\end{eqnarray}
the imaginary parts of the self-energies can be readily found to
be
\begin{eqnarray}
\mathrm{Im}\widetilde{\Sigma}^\mathrm{ret}_{\delta}(k,k_0) & =&
-\pi N_c N_f G^2 (k^2_0-k^2-4m^2) \int \frac{d^3p}{(2\pi)^3
2\omega_p \omega_q}\{[n(\omega_p)-n(\omega_q)]
[\delta(k_0-\omega_q+\omega_p)\nonumber\\
& & -\delta(k_0+\omega_q-\omega_p)]+[1-n(\omega_p)-n(\omega_q)]
[\delta(k_0+\omega_q+\omega_p)-
\delta(k_0-\omega_q-\omega_p)]\},\nn\\
\mathrm{Im}\widetilde{\Sigma}^{\mathrm{ret},ij}_{\pi}(k,k_0) & =&
-\pi \delta^{ij} N_c  N_f G^2 (k^2_0-k^2) \int
\frac{d^3p}{(2\pi)^3 2\omega_p\omega_q}\{[n(\omega_p)-n(\omega_q)]
[\delta(k_0-\omega_q+\omega_p)\nn\\
& & -\delta(k_0+\omega_q-\omega_p)]+[1-n(\omega_p)-n(\omega_q)
][\delta(k_0+\omega_q+\omega_p)-
\delta(k_0-\omega_q-\omega_p)]\}.\label{ImSigma}
\end{eqnarray}
The terms proportional to $[n(\omega_p)-n(\omega_q)]$ originate in
the process of Landau damping in which a meson scatters off a
quark in the medium, and those proportional to
$[1-n(\omega_p)-n(\omega_q) ]$ arise from processes in which quark
and antiquark are created via meson decay.

It is more convenient to obtain the Laplace transform
$\Sigma(k,s)$ for both self-energies from which the Matsubara and
real-time ones can be obtained by simple analytic continuations.
Furthermore we here focus on a long-wavelength, low-energy
expansion in the GL region for which $k,s,m \ll T\sim T_c$ and
keep terms up to $\mathcal{O}(k^2,s^2)$. After some algebra we
obtain
\begin{eqnarray}
\widetilde{\Sigma}_\delta(k,s)&=&N_c N_f G^2
\bigg\{(s^2+k^2+4m^2)\intL\frac{dp
p^2}{\pi^2\omega_p^2}\bigg[\frac{1-2\,n(\omega_p)}{4\omega_p}
+\frac{1}{2}\frac{dn(\omega_p)}{d\omega_p}\bigg(1-\frac{is\omega_p}{2kp}
\ln\frac{is/k+p/\omega_p}{is/k-p/\omega_p}\bigg)\bigg]\nn\\
&& -\intL\frac{dp p^2}{\pi^2}\frac{1-2\,n(\omega_p)}{\omega_p}+k^2
\intL \frac{dp}{12\pi^2} \frac{\partial}{\partial
p}\bigg[\frac{p^3}{\omega_p^3}[1-2\,n(\omega_p)]\bigg]\bigg\},\nn\\
\widetilde{\Sigma}^{ij}_\pi(k,s)& = & \delta^{ij} N_c N_f G^2
\bigg\{(s^2+k^2)\intL\frac{dp
p^2}{\pi^2\omega_p^2}\bigg[\frac{1-2n(\omega_p)}{4\omega_p}
+\frac{1}{2}\frac{dn(\omega_p)}{d\omega_p}\bigg(1-\frac{is\,\omega_p}{2kp}
\ln\frac{is/k+p/\omega_p}{is/k-p/\omega_p}\bigg)\bigg]\nn\\
&& -\intL\frac{dp p^2}{\pi^2}\frac{1-2\,n(\omega_p)}{\omega_p}+k^2
\intL \frac{dp}{12\pi^2} \frac{\partial}{\partial
p}\bigg[\frac{p^3}{\omega_p^3}[1-2\,n(\omega_p)]\bigg]\bigg\}.
\label{sigmalwsf}
\end{eqnarray}
We note that the last term in the above expressions is a surface
term which arises from the introduction of the three-momentum
cutoff $\Lambda$ that explicit breaks Lorentz invariance even at
zero temperature. If a covariant regularization scheme (e.g.,
dimensional regularization) is used, this surface term would
vanish identically. The necessity of neglecting this surface term,
a consequence of a non covariant regulator can be seen in the
zero-temperature case. After the analytic continuation
$is\rightarrow \omega +i\epsilon$ discarding this surface term
leads to the Lorentz invariant dispersion relation for scalar and
pseudoscalar fluctuations, namely $\omega^2 = k^2+4m^2$ for the
scalar ($\delta$) and $\omega^2=k^2$ for the pseudoscalar ($\pi$).
Hence the pions are the Goldstone bosons of the spontaneously
broken chiral symmetry. Obviously the surface term contains an
extra factor $k^2$ that spoils Lorentz invariance in the
dispersion relation at zero temperature. Therefore in order to
preserve the manifest Lorentz covariance, we will discard this
surface term hereafter.

The logarithmic dependence on $is/k$ in eq.~(\ref{sigmalwsf}) is a
consequence of the Landau damping contribution. Upon the analytic
continuation $is \rightarrow \omega+i\epsilon$ this term leads to
an imaginary part (see below) and a \emph{non-local} contribution
to the self-energy.

\subsection{The static effective action and Ginzburg-Landau effective theory}

The static limit corresponds to $s\rightarrow 0$ (or equivalently
$\nu_n=0$ in the Matsubara representation). The static limit of
the self-energies is given by
\begin{eqnarray}
\widetilde{\Sigma}_\delta(k,0)& = &
{\Sigma}_\delta(0,0)+\mathcal{C}(T)\,k^2,\nn \\
\widetilde{\Sigma}_\pi(k,0)& = &
{\Sigma}_\pi(0,0)+\mathcal{C}(T)\,k^2,\label{statdel}
\end{eqnarray}
where, upon neglecting the surface term in(\ref{sigmalwsf}), we
find the coefficient $\mathcal{C}(T)$ to be given by
\begin{equation}
\mathcal{C}(T)= \intL\frac{dp
p^2}{\pi^2\omega_p^2}\bigg[\frac{1-2n(\omega_p)}{4\omega_p}
+\frac{1}{2}\frac{d n(\omega_p)}{d\omega_p}\bigg].
\end{equation}
The long-wavelength limit of the static self-energies is given by
\begin{equation}
\Sigma_{\delta,\pi}(0,0)=\lim_{k\to 0}\lim_{s\to 0}
\widetilde{\Sigma}_{\delta,\pi}(k,s),
\end{equation}
which are found to be given by
\begin{eqnarray}
{\Sigma}_\delta(0,0) &=& -N_c N_f G^2 \intL\frac{dp
p^2}{\pi^2}\bigg[\bigg(1-\frac{m^2}{\omega_p^2}\bigg)
\frac{1-2n(\omega_p)}{\omega_p} -\frac{2m^2}{\omega_p^2}\frac{d
n(\omega_p)}{d\omega_p}\bigg],\nn\\
{\Sigma}_\pi(0,0) &=& -N_c N_f G^2 \intL\frac{dp
p^2}{\pi^2}\frac{1-2n(\omega_p)}{\omega_p}.
\end{eqnarray}

Comparing these expressions for the long-wavelength limit of the
static self-energies with the derivative of the effective
potential given by eq.~(\ref{gapeqn}), we find the following
identities
\begin{eqnarray}
1+\Sigma_{\delta}(0,0) & = &  \frac{d^2
\mathcal{V}_\mathrm{eff}(\sigma_0)}{d\sigma^2_0},\nn\\
1+\Sigma_{\pi}(0,0) & = &   \frac{1}{\sigma_0}\frac{d
\mathcal{V}_\mathrm{eff}(\sigma_0)}{d\sigma_0}.\label{pionmass}
\end{eqnarray}
Thus the \emph{static} effective action, obtained from
(\ref{Seff}) by taking the $\nu_n=0$ term, is given by
\begin{eqnarray}\label{Seffstat}
\frac{S_\mathrm{eff}[\delta,\pi;\sigma_0]}{\beta V}& =&
\mathcal{V}_\mathrm{eff}(\sigma_0)+ \delta
(0,\mathbf{0})\frac{d\mathcal{V}_\mathrm{eff}(\sigma_0)}{d\sigma_0}
+ \int \frac{d^3k}{(2\pi)^3}\nonumber \bigg\{\frac{1}{2}
\delta(0,\bk)\left[\frac{d^2
\mathcal{V}_\mathrm{eff}(\sigma_0)}{d\sigma^2_0}+\mathcal{C}(T)\,k^2
\right]\delta(0,-\bk)\nonumber \\ & &+ \frac{1}{2}
\pi^i(0,\bk)\left[\frac{1}{\sigma_0}\frac{d
\mathcal{V}_\mathrm{eff}(\sigma_0)}{\sigma_0}+\mathcal{C}(T)\,k^2
\right]\pi^i(0,-\bk)\bigg\},
\end{eqnarray}
with the effective potential in the GL region given by
\begin{equation}
\mathcal{V}_\mathrm{eff}(\sigma_0)= \frac{a}{2}(T^2-T^2_c)~
\sigma^2_0+ \frac{b}{4}~ \sigma^4_0 +
\mathcal{O}(\sigma^6_0),\label{VeffLGf}
\end{equation}
where the coefficients $a,b>0$ can be read off from
eq.~(\ref{VeffLG}).

Thus we see that up to a redefinition of the meson fields by
renormalizing their amplitudes by the constant $\mathcal{C}(T)$,
the \emph{static} low-energy effective action in the GL region is
\emph{exactly} of the form obtained from the linear sigma model
(\ref{NLSMF}) by assuming time-independent configurations, namely
setting the time derivatives in the kinetic terms in (\ref{NLSMF})
to zero.

The order parameter at the equilibrium minimum of the effective
potential for $T \lesssim T_c$ is given by the mean field solution
near the critical temperature
\begin{equation}
\sigma_\mathrm{eq}(T) = \left[\frac{a}{b}
(T^2_c-T^2)\right]^{\frac{1}{2}}\propto
(T_c-T)^{\frac{1}{2}},\label{MFS}
\end{equation}
where we have taken the positive solution. For fluctuations around
the equilibrium mean field solution in the chirally broken phase
for $T< T_c$, the multiplet of pions is massless as befits
Goldstone modes, whereas the fluctuations in the broken symmetry
direction have a mass determined by the second derivative of the
effective potential at the minimum.

This static effective action is equivalent to the expansion around
a mean field up to quadratic fluctuations of the O(4) GL free
energy
\begin{equation}
\mathcal{F}_\mathrm{GL}[\sigma,\pi] = \int d^3x \bigg [
\frac{\mathcal{C}(T)}{2} \boldsymbol{\nabla} \sigma \cdot
\boldsymbol{\nabla}\sigma + \frac{\mathcal{C}(T)}{2}
\boldsymbol{\nabla} \boldsymbol{\pi} \cdot
\boldsymbol{\nabla}\boldsymbol{\pi}+\frac{a}{2}(T^2-T^2_c)~
(\sigma^2_0+\boldsymbol{\pi}^2) + \frac{b}{4}~
(\sigma^2_0+\boldsymbol{\pi}^2)^2+h.o.t. \bigg].\label{LGfree}
\end{equation}
\noindent where $h.o.t.$ stands for higher order terms in
$\sigma^2_0,\mathbf{\pi}^2$ and gradients. The equivalence between
(\ref{LGfree}) and (\ref{VeffLGf}) up to quadratic order in the
fluctuations $\delta, \pi$ can be immediately seen by writing
$\sigma=\sigma_0+\delta$ and keeping only up to quadratic terms in
$\delta, \pi$ in the free energy (\ref{LGfree}), in agreement with
the discussion leading to eq.~(\ref{NLSMF}). Thus we see that
long-wavelength phenomena in the \emph{static limit} is in the
same universality class as the \emph{three-dimensional} O(4)
Heisenberg ferromagnet (linear sigma model) as advanced in
Ref.~\cite{pisarski,wilraja}.

The local gradient expansion that allows the identification of the
static effective action with the GL free energy of the O(4)
Heisenberg ferromagnet, is available only in the static limit. The
analytic continuation of the self-energies to real frequencies are
both non-local and complex. We now address the issue of the
dynamics.

\section{Real-time dynamics}\label{sec:realtime}

The real-time evolution of expectation values and correlation
functions is obtained from the real-time generating functional
\begin{equation}
\mathcal{Z}[J^{\pm}] =
\mathrm{Tr}U(\infty,-\infty;J^+)\,\rho\,U^{-1}(\infty,-\infty;J^-),
\label{RTZ}
\end{equation}
where $\rho$ is an initial density matrix and $U(t,t';J^+)$
[$U^{-1}(t,t';J^-)$] is the forward (backwards) time evolution
operator in presence of a current $J^+$ ($J^-$). Time ordered,
anti-time ordered, advanced and retarded correlation functions are
obtained by taking suitable variational derivatives with respect
to the sources $J^\pm$. In the case under consideration, we seek
to study the real-time evolution of small amplitude fluctuations
around a mean field $\sigma_0$, hence we take the initial density
matrix to be thermal at temperature $T<T_c$ in the broken symmetry
phase.

We will study the real-time evolution of small amplitude
fluctuations as an initial value problem, which is defined by
introducing \emph{external} sources $\eta_{\delta,\pi}$ that
induce an expectation value for the fluctuations around the mean
field. The real-time generating functional (\ref{RTZ}) can be
conveniently written in a path integral representation as follows
\begin{equation}\label{parfun}
\mathcal{Z}[J^{\pm};\eta] = \int
\mathcal{D}\bar{\psi}^\pm\mathcal{D}{\psi^\pm}\mathcal{D}
{\delta^\pm}\mathcal{D}{\boldsymbol{\pi}^\pm}\,e^{i\int_{-\infty}^{\infty}
d^4x (\mathcal{L}^+[J^+,\eta] - \mathcal{L}^-[J^-,\eta])},
\end{equation}
with
\begin{eqnarray}
\mathcal{L}^\pm[J^\pm,\eta]  & = &
\bar{\psi}^{\pm}(\bx,t)\left[i\!\!\not\!\partial-m-G
\delta^{\pm}(\bx,t)-i G \gamma^5 \boldsymbol{\tau}\cdot
\boldsymbol{\pi}^{\pm}(\bx,t)\right]\psi^{\pm}(\bx,t)+
\bar{J}^\pm_{\psi}(\bx,t) \psi^\pm(\bx,t)\nonumber \\
&&+\bar{\psi}^\pm(\bx,t)J^\pm_{\psi}(\bx,t)-\sigma_0
\delta^\pm(\bx,t)-\frac{1}{2} \delta^{2,\pm}(\bx,t)-\frac{1}{2}
\boldsymbol{\pi}^{2,\pm}(\bx,t)+J^\pm_{\delta}(\bx,t)\delta^\pm(\bx,t)\nn\\
&&+\mathbf{J}^\pm_{\pi}(\bx,t)\cdot\boldsymbol{\pi}^\pm(\bx,t)+
\eta_{\delta}(\bx,t)\delta^\pm(\bx,t)+
\boldsymbol{\eta}_{\pi}(\bx,t)\cdot\boldsymbol{\pi}^\pm(\bx,t),
\label{RTLagra}
\end{eqnarray}
where the superscripts $\pm$ refer to fields defined in the
forward ($+$) and backward ($-$) branches of the closed-time-path
contour in the Schwinger-Keldysh formalism of nonequilibrium field
theory~\cite{ctp}. In the Lagrangian densities above, the sources
$J^\pm$ are auxiliary, real-time correlation functions are
obtained as variational derivatives with respect to these and
setting them to zero afterwards. In contrast, the external sources
$\eta_{\delta,\pi}$ above are introduced to induce an expectation
value for the fluctuations $\delta,\pi$ and allow to study the
real-time evolution of these expectation values as an
\emph{initial value problem} as described in detail below.

Perturbative calculations for the meson self-energies are carried
out in terms of the following real-time free fermion propagators:
\begin{eqnarray}
&&\langle \psi^{a}(\bx,t) \bar{\psi}^{b}({\bx}',t')\rangle = i
\intp S_\bp^{ab}(t,t')\, e^{i\bp\cdot({\bx}-{\bx}')} \; ,\nn\\
&&S_\bp^{++}(t,t')= S_\bp^>(t,t') \; \Theta(t-t') +S_\bp^<(t,t')
\; \Theta(t'-t), \nn\\ &&S_\bp^{--}(t,t')= S_\bp^>(t,t') \;
\Theta(t'-t)+S_\bp^<(t,t') \; \Theta(t-t') \; ,\nn\\
&&S_\bp^{\pm\mp}(t,t')= S_\bp^\lessgtr(t,t') \; ,
\label{fermionprop1}
\end{eqnarray}
where $\langle\cdots\rangle$ denotes expectation value with
respect to the initial density matrix and $a, b=\pm$. The Wightman
functions $S_\bp^\lessgtr(t,t')$ are expressed in terms of the
spectral function as (with zero chemical potential)
\begin{eqnarray}
S_\bp^>(t,t')&=& -i\int^\infty_{-\infty}dp_0\,
\rho(\bp,p_0)\,[1-n(p_0)]\,e^{-ip_0(t-t^\prime)},\nn\\
S_\bp^<(t,t')&=& i\int^\infty_{-\infty}dp_0\, \rho(\bp,p_0)\,
n(p_0)\, e^{-ip_0(t-t^\prime)},\label{wightmanS}
\end{eqnarray}
where $n(p_0)= 1/(e^{\beta p_0}+1)$ and  the spectral function
$\rho(\bp,p_0)$ is given by eq.~(\ref{propDR}).

\subsection{Linear relaxation of  fluctuations}

To study the real-time dynamical evolution of the small amplitude
perturbation of the $\sigma$ and $\pi$ fields, we consider
preparing a system in the broken symmetry phase $\sigma_0\neq 0$,
slightly perturbed away from equilibrium by adiabatically
switching-on the external sources $\eta_{\delta,\pi}$ in the
infinite past (see below). Once the sources are switched-off at
some time, say $t=0$, the perturbed system relaxes towards
equilibrium and the relaxation dynamics is studied in linear
response. Since the external sources $\eta_{\delta,\pi}$ induce
expectation values for $\delta$ and $\pi$, it is convenient to
decompose $\delta$ and $\pi$ into an expectation value and a
fluctuation whose expectation value vanishes (even with
$\eta_{\delta,\pi}$ switched-on)~\cite{ivp}
\begin{equation}
\delta^\pm(\bx,t)=\phi(\bx,t)+\Delta^\pm(\bx,t),
\quad\langle\Delta^\pm(\bx,t)\rangle_{\eta}=0.
\end{equation}
Likewise, the $\pi$ field is decomposed as
\begin{equation}
\pi^{i,\pm}(\bx,t)=\chi^i(\bx,t)+\Pi^{i,\pm}(\bx,t),\quad
\langle\Pi^{i,\pm}(\bx,t)\rangle_{\eta}=0,
\end{equation}
where $\chi^i(\bx,t)$ is the fluctuation about the homogeneous
pseudoscalar condensate which we have chosen to vanish. The
expectation values $\langle \Pi^\pm(\bx,t)\rangle_{\eta}$ and
$\langle \Delta^\pm(\bx,t)\rangle_{\eta}$  are obtained
consistently in a loop expansion with the real-time generating
functional in presence of the external sources $\eta$.

The equations of motion for the expectation values in linear
response are obtained implementing the tadpole conditions $\langle
\Pi^\pm(\bx,t)\rangle_{\eta}=0$ and $\langle
\Delta^\pm(\bx,t)\rangle_{\eta}=0$. These expectation values are
obtained in a systematic loop expansion in terms of quark loops
but only to linear order in $\phi(\bx,t);\chi^i(\bx,t)$ consistent
with linear response. This method has been implemented in many
different problems, and the reader is referred to
reference\cite{tadpole} for more details.

The \emph{linearized} equations of motion for the spatial Fourier
transform of the small amplitude condensate fluctuations
$\phi(\xt)$ and $\chi(\xt)$ are the following
\begin{eqnarray}
&&\phi(\kt)+\int^{t}_{-\infty}dt'\Sigma_\delta(k,t-t')\phi(\kt')
+\frac{d\mathcal{V}_\mathrm{eff}(\sigma_0)}{d\sigma_0}\delta^{(3)}(\bk)
=\,\eta_\delta(\kt),\nn\\
&&\chi^i(\kt)+\int^{t}_{-\infty}dt'\Sigma^{ij}_\pi(k,t-t')\chi^j(\kt')=
\eta^i_\pi(\kt), \label{eom1}
\end{eqnarray}
respectively. Since the quark fields are described by a thermal
initial density matrix, the real-time self-energies
$\Sigma_{\delta,\pi}(k,t-t')$ are related to the respective
retarded self-energies of the $\sigma$ and $\pi$ fields given by
eqs.~(\ref{retSE})-(\ref{disret}) and \eqref{ImSigma}. In this
section we  consider fluctuations around the equilibrium state,
namely $d\mathcal{V}_\mathrm{eff}(\sigma_0)/d\sigma_0=0$,
postponing the discussion of fluctuations off the equilibrium
minimum to a later section (see sec. \ref{sec:spino}).

The adiabatic preparation of a system slightly perturbed away from
equilibrium can be realized within the real-time formulation by
taking the spatial Fourier transform of the external sources to be
of the form~\cite{ivp}
\begin{equation}
\eta_{\delta,\pi}(\bk,t)=\eta_{\delta,\pi}(\bk)\,e^{\epsilon
t}\,\Theta(-t), \quad \epsilon \to 0^+.\label{adsource}
\end{equation}
The $\epsilon$-term serves to switch on the source adiabatically
from $t=-\infty$ so as not to disturb the system too far from
equilibrium in the process. If the system was in an equilibrium
state at $t=-\infty$, then the equilibrium condition
$d\mathcal{V}_\mathrm{eff}(\sigma_0)/d\sigma_0=0$ ensures that for
$t<0$ there is a solution of the equations of motion (\ref{eom1})
of the form
\begin{equation}
\left\{
\begin{array}{l}
\phi(\bk,t) = \phi(\bk,0)\,e^{\epsilon t}\\
\chi^i(\bk,t) = \chi^i(\bk,0)\,e^{\epsilon t}
\end{array}\right.
\quad\mathrm{for}\;\;t<0,\label{solless0}
\end{equation}
where $\phi(\bk,0)$ [$\chi^i(\bk,0)$] is related to
$\eta_\delta(\bk)$ [$\eta^i_\pi(\bk)$] through the equations of
motion for $t<0$. That this is a consistent solution for $t<0$ can
be seen easily from the representation (\ref{intrep}).

The integrals over $t'$ in the non-local terms in the equations of
motion (\ref{eom1}) can be separated into $t'<0$ and $t'>0$, in
the interval $-\infty< t'<0$ the solution is given by
eq.~(\ref{solless0}). There only remains a convolution for $t'>0$
which is studied by taking the Laplace transform of the equations
of motion for $t>0$. The equations of motion for the fluctuations,
in terms of the Laplace variable $s$, are given by
\begin{eqnarray}
\widetilde{D}^{-1}_\delta(k,s)\widetilde{\phi}(\bk,s)&=&
\frac{1}{s}\left[\widetilde{D}^{-1}_\delta(k,s)-
\widetilde{D}^{-1}_\delta(k,0)\right]
\phi(\bk,0),\nn\\
\widetilde{D}^{-1}_\pi(k,s)\widetilde{\chi}^i(\bk,s)&=&
\frac{1}{s}\left[\widetilde{D}^{-1}_\pi(k,s)-
\widetilde{D}^{-1}_\pi(k,0)\right]\chi^i(\bk,0),\label{eomlaplace}
\end{eqnarray}
where $\widetilde{\phi}(\bk,s)$ [$\widetilde{\chi}^i(\bk,s)$] is
the Laplace transform of $\phi(\bk,t)$ [$\chi(\bk,t)$]. In the
above equations, $\widetilde{D}^{-1}_{\delta,\pi}(k,s)$ is the
inverse retarded propagator of the corresponding field, namely
\begin{equation}
\widetilde{D}^{-1}_{\delta,\pi}(k,s)= 1+
\widetilde{\Sigma}_{\delta,\pi}(k,s),
\end{equation}
where $\widetilde{\Sigma}_{\delta,\pi}(k,s)$ is the Laplace
transform of the self-energy given by equation (\ref{LaplaSig}).
In the long-wavelength, low-energy limit, these Laplace transforms
are given by equation (\ref{sigmalwsf}). The self-energy
$\Sigma_{\delta,\pi}(k,\omega)$ and inverse propagator
$D^{-1}_{\delta,\pi}(k,\omega)$ in term of the real frequency
$\omega$ are defined as the boundary values of the analytic
functions $\widetilde{\Sigma}_{\delta,\pi}(k,s)$ and
$\widetilde{D}^{-1}_{\delta,\pi}(k,s)$, respectively, through the
analytic continuation $s\to i\omega+\epsilon$.

To simplify the notation, we will henceforth drop the isospin
indices of the pion field by invoking the remaining
$\mathrm{SU}(2)_\mathrm{V}$ isospin symmetry. The real-time
evolution of the small amplitude fluctuations $\phi(\kt)$ and
$\chi(\kt)$ for $t>0$ is now in the form of an \emph{initial value
problem} with the respective initial value specified at $t=0$
given by $\phi(\bk,0)$ and $\chi(\bk,0)$. The solution of this
initial value problem can be obtained from the inverse Laplace
transform, e.g., [similar expression for $\phi(\kt)$]
\begin{equation}
\chi(\kt)=\int_\mathcal{B}\frac{ds}{2\pi
i}\,e^{st}\,\widetilde{\chi}(\bk,s), \label{invlap}
\end{equation}
where $\widetilde{\chi}(\bk,s)$ is the solution of
\eqref{eomlaplace} and the Bromwich contour $\mathcal{B}$ runs
parallel to the imaginary axis in the complex $s$-plane to the
right of all the singularities (isolated poles and branch cuts) of
$\widetilde{\chi}(\bk,s)$~\cite{ivp}. We note that there is
\emph{no} isolated pole in $\widetilde{\phi}(\bk,s)$ and
$\widetilde{\chi}(\bk,s)$ at $s=0$ since the corresponding residue
vanishes.

We now focus our attention to the real-time dynamics of small
amplitude fluctuations near the equilibrium minimum determined by
$d\mathcal{V}_\mathrm{eff}(\sigma_0)/d\sigma_0 =0$. It proves
convenient to introduce the following dimensionless variables
\begin{equation}
\bar{k}=\frac{k}{T},\quad\bar{s}=\frac{s}{T},\quad, \quad
\xi=\frac{s}{k},\quad
x=\frac{m}{T},\quad\bar{\Lambda}=\frac{\Lambda}{T},\quad g=N_c N_f
G^2 T^2,
\end{equation}
the inverse propagators of the $\sigma$ and $\pi$ fields now read
\begin{eqnarray}
\widetilde{D}^{-1}_{\sigma}(k,s)&=&g(\bar{s}^2+\bar{k}^2+4x^2)
H(i\xi,x),\nn\\
\widetilde{D}^{-1}_{\pi}(k,s)&=&g(\bar{s}^2+\bar{k}^2)
H(i\xi,x),\label{Dinverse}
\end{eqnarray}
where we have used eq.~(\ref{gapeqn}) and introduced
\begin{equation}
H(i\xi,x)=\frac{1}{4\pi^2}\int_0^{\bar{\Lambda}}\frac{dz
 z^2}{\epsilon^2(z,x)}\bigg[\frac{\tanh\frac{\epsilon(z,x)}{2}}{\epsilon(z,x)}-
\frac{1}{2\cosh^{2}\frac{\epsilon(z,x)}{2}}
\bigg(1-\frac{i\xi\epsilon(z,x)}{2z}
\ln\frac{i\xi+z/\epsilon(z,x)}{i\xi-z/\epsilon(z,x)}\bigg)\bigg],~~;~~
\epsilon(z,x)=\sqrt{z^2+x^2} \label{H}
\end{equation}

The (quasi)particle excitations of the scalar meson are massive
and as discussed in Ref.~\cite{hatsuda} they correspond to a broad
resonance arising from the decay of this meson into
quark-antiquark pairs even at lowest (one-loop) order. Physically
the scalar excitations are not stable. In what follows we will
restrict our attention to the dynamics of pions which are the true
low-energy degrees of freedom in QCD.


While the one loop self-energies for the scalar and pseudoscalars
are fairly well known\cite{hatsuda,klevansky} and to this one loop
order, or alternatively leading order in the large $N_c N_f$
corresponds to the random phase approximation (summation of ring
diagrams for the propagators), unlike previous studies, we focus
on the \emph{real time dynamics} for which the formulation
described above is necessary. The analysis of the static case,
while in agreement with previous studies\cite{hatsuda,klevansky}
allows to study on the same footing and under the same
approximations both the static as well as dynamical aspects thus
allowing an unambiguous comparison. In particular the explicit
form of the static effective action obtained above will allow us
to compare the predictions for the growth rate of spinodally
unstable long-wavelength fluctuations as obtained from the static
effective action and those obtained from a detailed analysis of
the dynamic response.


\begin{figure}[t]
\includegraphics[width=3.5in,keepaspectratio=true]{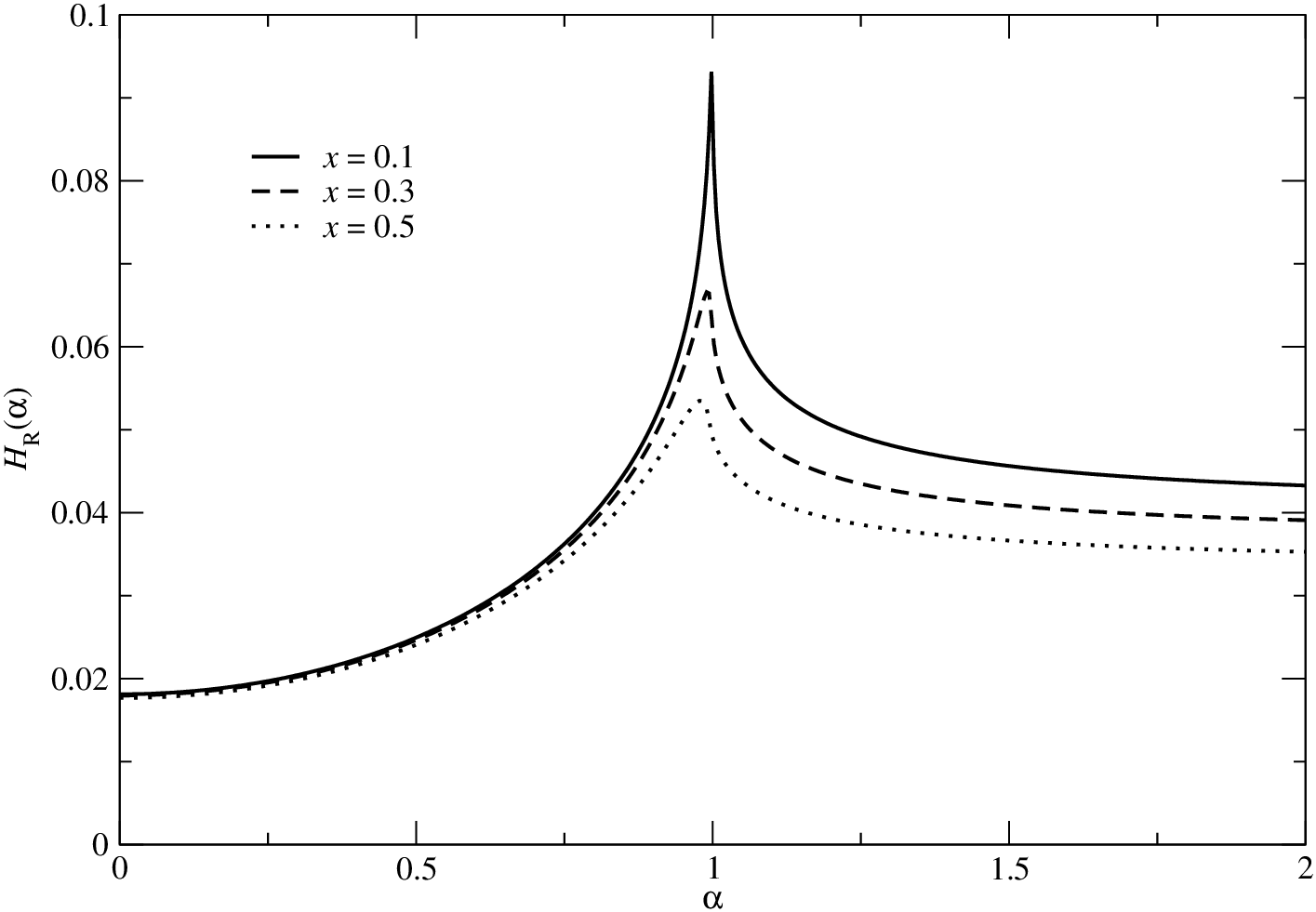}
\caption{The function $H_\mathrm{R}(\alpha,x)$ vs $\alpha$ for
different values of $x$.}\label{fig:HR}
\end{figure}

\subsection{Pion dynamics}

\subsubsection{Fluctuations around equilibrium}

We begin our study by focusing on the real-time (quadratic)
fluctuations around the equilibrium broken symmetry state for $T<
T_c$, which is determined by the condition $
d\mathcal{V}_\mathrm{eff}(\sigma_0)/d\sigma_0=0$ with $\sigma_0
\neq 0$. In this case, the real-time dynamics of pions is
determined by the Laplace transform
\begin{equation}
\widetilde{\chi}(\bk,s)=\frac{1}{s}\left[1-\frac{H(0)}{(1+\xi^2)
H(i\xi,x)}\right]\chi(\bk,0).
\end{equation}

After the analytic continuation $s\to i\omega+\epsilon$ and
introducing the dimensionless variable $\alpha=\omega/k$, we find
the real and imaginary parts of the analytically continued
function $H(\alpha+i\epsilon)$ given by
\begin{eqnarray}
H_\mathrm{R}(\alpha,x)&=&\frac{1}{4\pi^2}\int_0^{\bar{\Lambda}}\frac{dz
z^2}{\epsilon^2}\bigg[\frac{\tanh\frac{\epsilon(z,x)}{2}}{\epsilon(z,x)}-
\frac{1}{2\cosh^{2}\frac{\epsilon(z,x)}{2}}
\bigg(1-\frac{\alpha\epsilon(z,x)}{2z}
\ln\left|\frac{\alpha+z/\epsilon(z,x)}{\alpha-z/\epsilon(z,x)}\right|\bigg)\bigg],\nn\\
H_\mathrm{I}(\alpha,x)&=&\frac{\alpha}{16\pi}\,
\Theta(\alpha_\mathrm{max}^2-\alpha^2)
\int_{z_\mathrm{min}}^{\bar{\Lambda}}\frac{dz
z}{\epsilon(z,x)\cosh^{2}\frac{\epsilon(z,x)}{2}},\label{HRHI}
\end{eqnarray}
respectively,

\begin{equation}
\alpha_\mathrm{max}=\bar{\Lambda}/\sqrt{\bar{\Lambda}^2+x^2}~~;~~
z_\mathrm{min}=x\alpha/\sqrt{1-\alpha^2}\end{equation}

The real part can be written in the following manner which
separates the terms that are most sensitive to the cutoff

\begin{eqnarray}
H_\mathrm{R}(\alpha,x) & = & \frac{1}{4\pi^2}
\ln(\bar{\Lambda})+\frac{1}{4\pi^2} \int_0^{\bar{\Lambda}}
\frac{dz}{z} \Bigg[ \frac{\tanh\left(\frac{1}{2}\sqrt{x^2+z^2}
\right)}{\left(1+\frac{x^2}{z^2} \right)^{3/2}}-\Theta(z-1)
\Bigg]+B(\alpha,x) \label{firstterm} \\
B(\alpha,x) & = & - \frac{1}{4\pi^2}\int_0^{\bar{\Lambda}}\frac{dz
z^2}{\epsilon^2}\frac{1}{2\cosh^{2}\frac{\epsilon(z,x)}{2}}
\bigg(1-\frac{\alpha\epsilon(z,x)}{2z}
\ln\left|\frac{\alpha+z/\epsilon(z,x)}{\alpha-z/\epsilon(z,x)}\right|\bigg)\label{secterm}
\end{eqnarray}

The second term in eqn. (\ref{firstterm}) is rather insensitive to
$\bar{\Lambda}$ and in this expression $\bar{\Lambda}$ can be
taken to infinity with an exponentially small error. The form of
$H(\alpha,x)$ above displays explicitly the dependence on
$\bar{\Lambda}$.

\begin{figure}[t]
\includegraphics[width=3.5in,keepaspectratio=true]{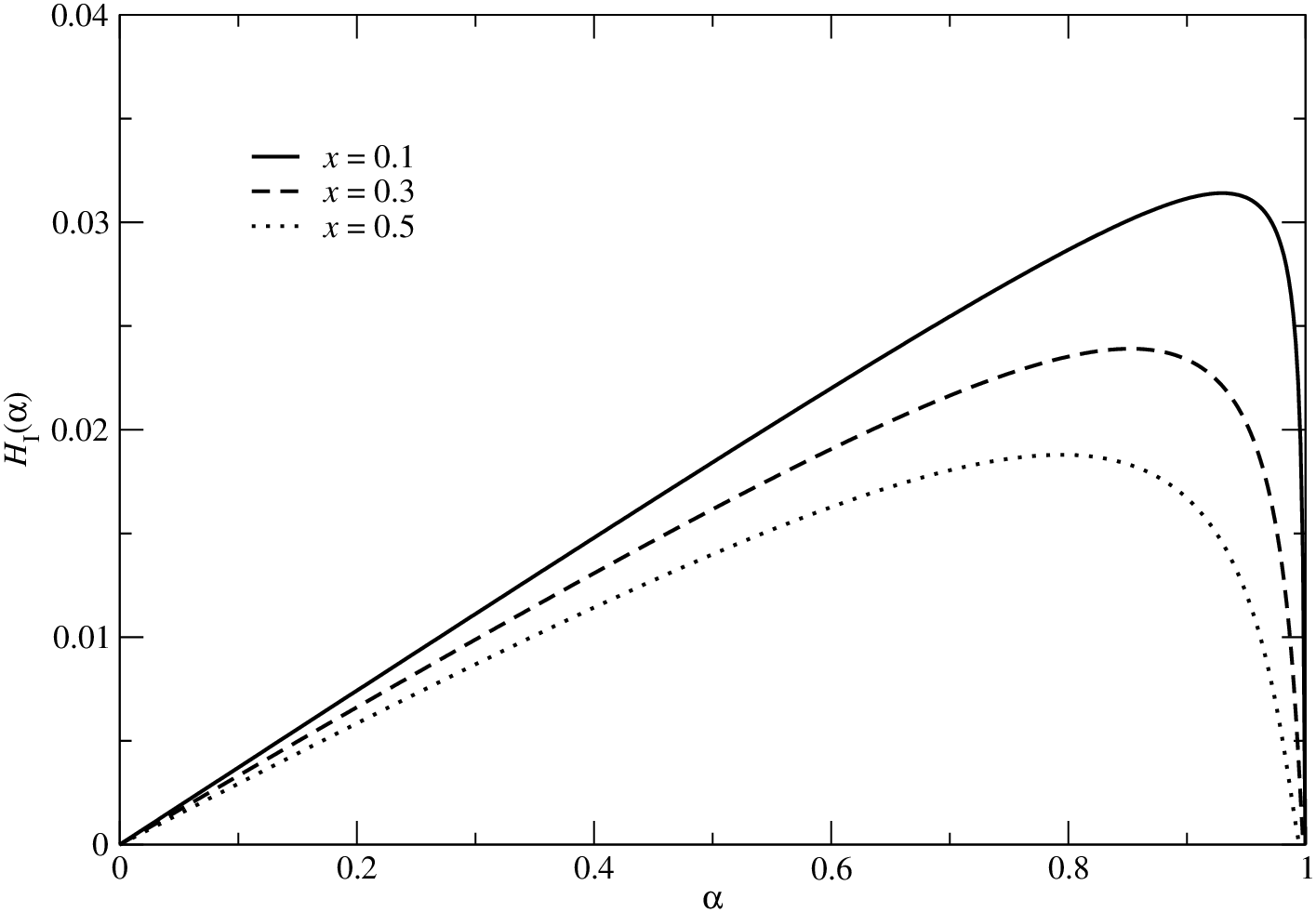}
\caption{The function $H_\mathrm{I}(\alpha,x)$ vs $\alpha$ for
different values of $x$.}\label{fig:HI}
\end{figure}

The real and imaginary parts given in \eqref{HRHI} are plotted in
Figs.~\ref{fig:HR} and \ref{fig:HI}, respectively, for
$N_c=3,N_f=2$, the values of $G,\Lambda$ given by eqn.
(\ref{parameters}) and different values of $x$  in the
Landau-Ginzburg region $x \ll 1$.

 The real part $H_\mathrm{R}(\alpha,x)$ increases
monotonically for $0<\alpha\lesssim \alpha_\mathrm{max}$ then
decreases monotonically to its asymptotic value for $\alpha\gg 1$.
Furthermore, $H_\mathrm{R}(\alpha,x)$ features a sharp cusp near
$\alpha\lesssim\alpha_\mathrm{max}$ as $x\to 0$. This cusp implies
infrared divergence in the self-energies at the critical point
$x=0$ and, as will be seen below, has a importance consequence on
the relaxation of critical fluctuations. The imaginary part
$H_\mathrm{I}(\alpha,x)$, which is completely determined by Landau
damping only has support in the interval
$-\alpha_\mathrm{max}<\alpha<\alpha_\mathrm{max}$.

The analytic properties of the quark loop diagram at finite
temperature had also been studied in ref.\cite{caldas}.

The analytic continuation of the propagator to real frequency
features the following singularities:
\begin{itemize}
\item{\textbf{Isolated poles}. Isolated \emph{real} poles describe
stable quasiparticle excitations, whereas  \emph{complex} poles
(on the second sheet) correspond to resonances with finite width.
From (\ref{eomlaplace}) and (\ref{Dinverse}), isolated poles of
the Laplace transform for the pion  field are determined by the
equations:
\begin{equation}
(1-\alpha^2)H_\mathrm{R}(\alpha,x) = 0,\quad
(1-\alpha^2)H_\mathrm{I}(\alpha,x)= 0.\label{isopoles}
\end{equation}
Since $H_\mathrm{I}(\alpha,x)$ vanishes at $\alpha=\pm 1$, there
are isolated real poles at $\alpha=\pm 1$ provided that
$H_\mathrm{R}(\alpha,x)$ remains finite there. This indeed is the
case below the critical point when $x>0$. Specifically,
$H_\mathrm{R}(1)$ is a monotonically decreasing function for $x>0$
and its asymptotic behavior for $x\ll 1$ is found to be given by
\begin{equation}\label{H(1)}
H_\mathrm{R}(1)\stackrel{x\ll 1}{\simeq}-\frac{1}{4\pi^2}\Bigg[
\ln\left(\frac{x}{\pi}\right)+\mathcal{O}(x^0)\Bigg].
\end{equation}
Thus for $x>0$ pions are sharp quasiparticles in this
approximation that propagate with a dispersion relation
$\omega_{\pi}(k)= \pm k$, namely with unit group velocity. }

\item{\textbf{Branch cuts}. Branch cuts represent multiparticle
states. In the long-wavelength, low-frequency limit there is only
a branch cut on the real frequency axis in the region
$-\alpha_\mathrm{max}k <\omega<\alpha_\mathrm{max}k$,
corresponding to Landau damping processes in which the pion
scatters off a quark in the medium.}
\end{itemize}
Having identified the singularities of the Laplace transform
$\widetilde{\chi}(\bk,s)$, we can perform the inverse Laplace
transform (\ref{invlap}) and obtain the real-time evolution
\begin{eqnarray}
\frac{\chi(\bk,t)}{\chi(\bk,0)}=\int_{-\infty}^\infty
\frac{d\alpha}{\alpha}\rho(\alpha)\cos\alpha\tau,\label{chit}
\end{eqnarray}
where $\tau=k t$, $\rho(\alpha)$ is the spectral function for the
pion. In what follows we  set $\chi(\bk,0)=1$ for notational
simplicity. The spectral function $\rho(\alpha)$ receives
contributions from the quasiparticle poles and the Landau damping
cut:
\begin{equation}
\rho(\alpha)=
Z\,\mathrm{sgn}(\alpha)\,\delta(1-\alpha^2)+\rho_\mathrm{cut}(\alpha),
\end{equation}
where $\mathrm{sgn}(\alpha)$ is the sign function, $Z$ is the
residue of the pion poles at $\alpha=\pm 1$, and
$\rho_\mathrm{cut}(\alpha)$ denotes the cut contribution
\begin{equation}
Z=\frac{H_\mathrm{R}(0,x)}{H_\mathrm{R}(1,x)},\quad
\rho_\mathrm{cut}(\alpha)=
\frac{H_\mathrm{R}(0,x)H_\mathrm{I}(\alpha,x)}{\pi(1-\alpha^2)
[H_\mathrm{R}^2(\alpha,x)+ H_\mathrm{I}^2(\alpha,x)]}.
\end{equation}
Therefore the real-time solution is given by
\begin{equation}
\chi(\bk,t)=\chi_\mathrm{pole}(\bk,t)+\chi_\mathrm{cut}(\bk,t),
\end{equation}
where for $\chi(\bk,0)=1$, we find
\begin{eqnarray}
&&\chi_\mathrm{pole}(\bk,t) =  Z\cos\tau,\nonumber \\
&&\chi_\mathrm{cut}(\bk,t)  =  2\int_0^{\alpha_\mathrm{max}}
\frac{d\alpha}{\alpha}\rho_\mathrm{cut}(\alpha)\cos\alpha\tau,
\quad \tau = k t. \label{realtimechi}
\end{eqnarray}
Setting $t=0$ in (\ref{realtimechi}) [with $\chi(\bk,0)=1$], we
obtain the following sum rule for the spectral function
\begin{equation}
Z+2\int_0^{\alpha_\mathrm{max}}
\frac{d\alpha}{\alpha}\rho_\mathrm{cut}(\alpha)=1,
\end{equation}
which we have verified numerically for a wide range of $x$.

\begin{figure}[ht!]
\includegraphics[width=3.5in,height=2in]{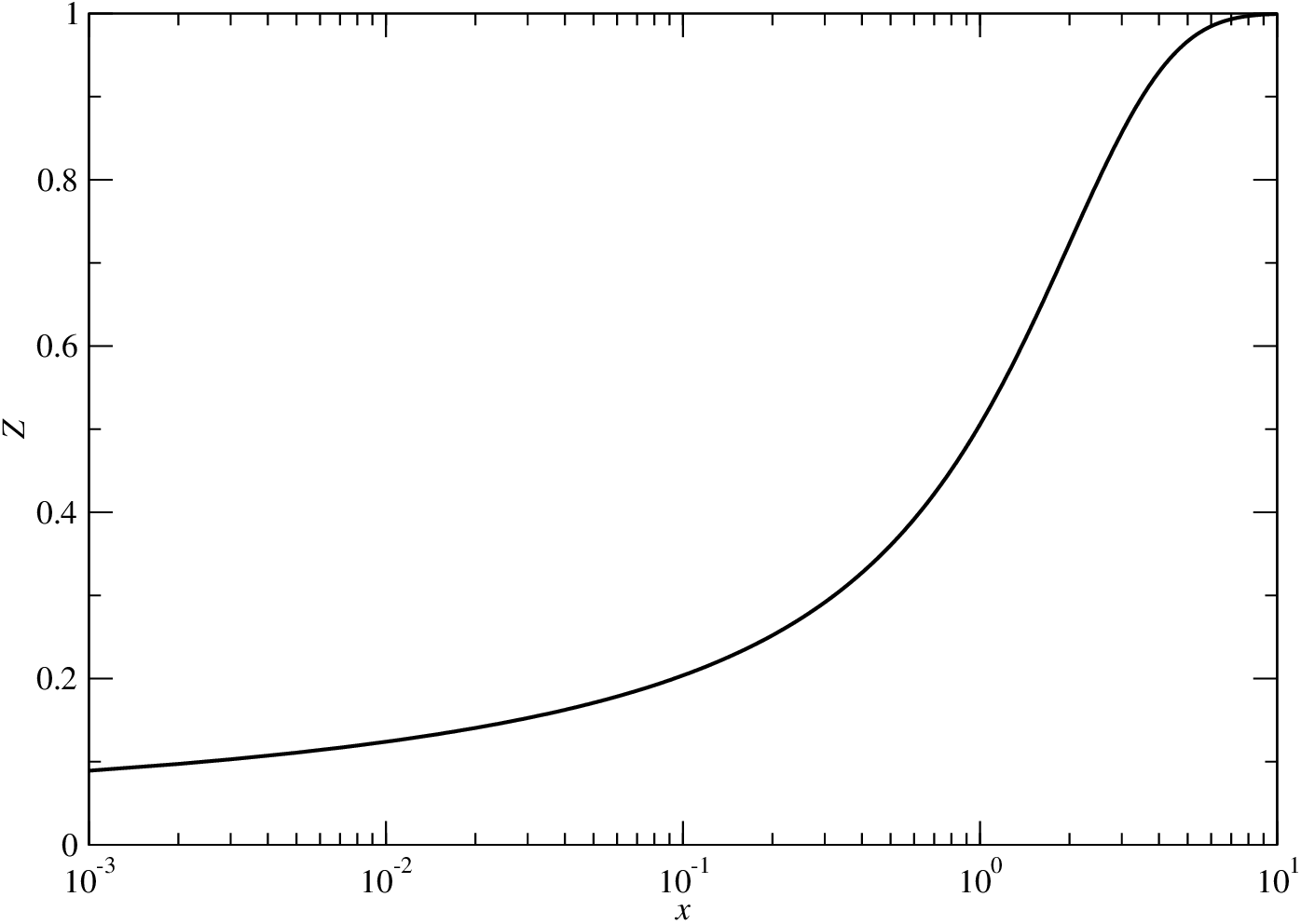}
\caption{The residue of the pion pole $Z$ vs $x$.}\label{fig:Z}
\end{figure}

\begin{figure}[h]
\includegraphics[width=3in,height=1.8in]{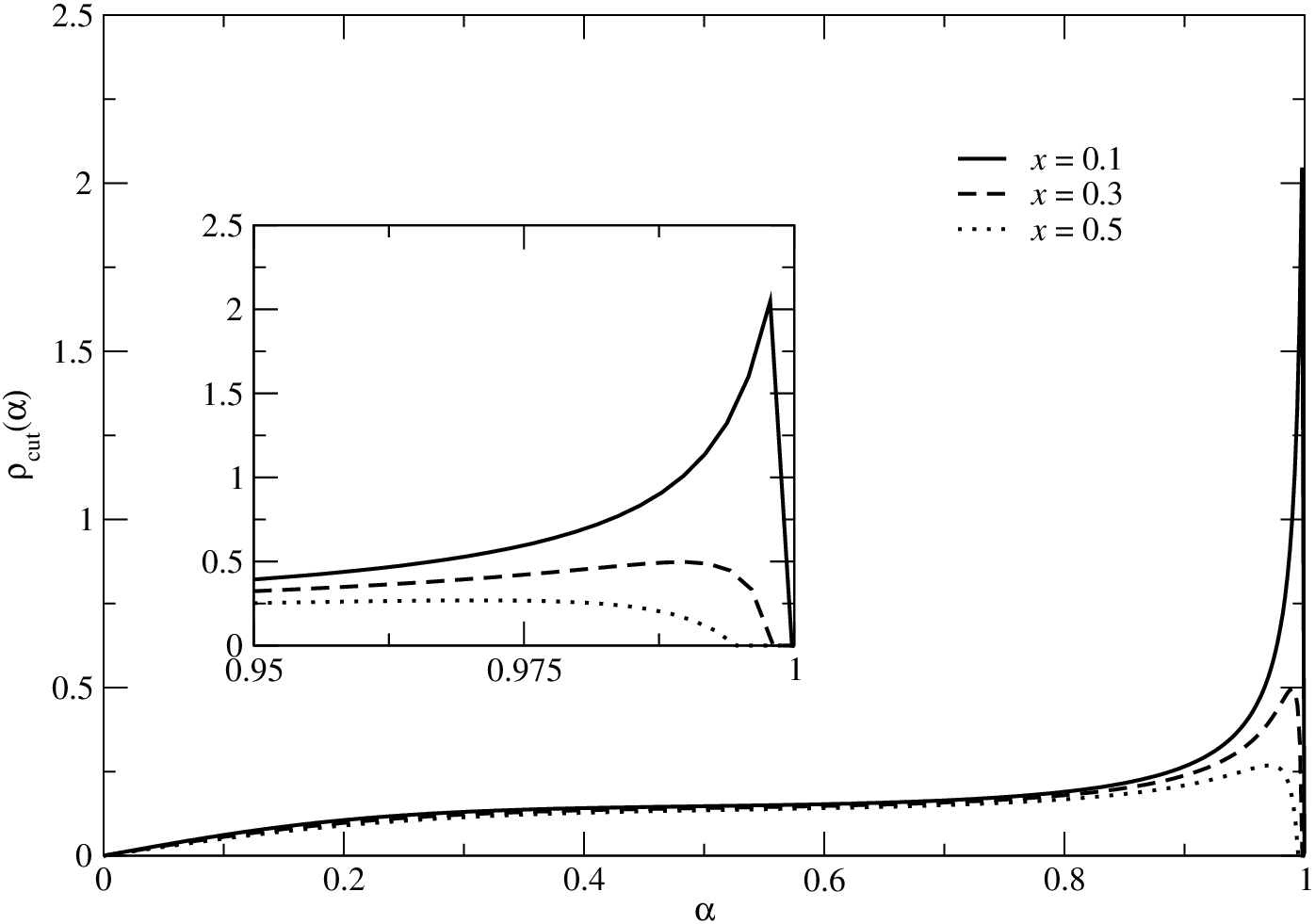}
\caption{The cut contribution to the spectral function
$\rho_\mathrm{cut}(\alpha)$ vs $\alpha$ for different values of
$x$. The inset shows the behavior of $\rho_\mathrm{cut}(\alpha)$
near the threshold
$\alpha=\alpha_\mathrm{max}$.}\label{fig:rhocut}
\end{figure}

The quasiparticle poles give rise to an undamped oscillation with
an amplitude determined by the residue $Z$. Figure~\ref{fig:Z}
displays the residue of the quasiparticle pole $Z$ as a function
of $x$. It shows clearly that away from the GL region in the
low-temperature limit ($x \gg 1$) the spectral function is
dominated by the isolated pion pole since $Z \approx 1$. Thus
Landau damping is negligible and pions are stable excitations.
This is because Landau damping is solely a medium effect arising
from scatterings of pions with quarks in the heat bath, hence its
contribution at lower temperature is less important. In the GL
region ($x\ll 1$) the continuum contribution from the Landau
damping cut dominates the spectral function, since in this region
$Z \ll 1$. Figure~\ref{fig:Z} also shows a crossover at $x \sim 1$
at which $Z\sim 1/2$, hence the isolated pole and the continuum
contributions are of equal importance. The Landau damping
contribution to the spectral function $\rho_\mathrm{cut}(\alpha)$
is displayed in Fig.~\ref{fig:rhocut}. For $x\ll 1$, it reveals
clearly a sharp peak near the threshold, indicating the emergence
of an infrared divergence at the critical point.

Equation (\ref{H(1)}) and the fact that $H_\mathrm{R}(0)$ is
finite and non-vanishing at $x=0$ entail that near the critical
temperature, namely $x\ll 1$, the residue of the pion pole behaves
as
\begin{equation}
Z \propto \frac{1}{\left|\ln\frac{T_c-T}{T_c}\right|}.\label{ZTc}
\end{equation}

\noindent where we have used that $x=m(T)/T$ and $m(T) =
G\sigma_{eq}(T)\propto (T_c-T)^{1/2}$ by eqn. (\ref{MFS}).

Therefore the residue at the isolated pion pole vanishes as
$T\rightarrow T_c$ from below. As the critical point is
approached, the weight of the isolated pole to the spectral
function becomes negligible and exactly at the critical point the
residue vanishes and the isolated pole merges with the continuum
contribution. Thus it is clear that as the critical region is
approached, the edge of the Landau damping cut begins to dominate
the spectral function.

\begin{figure}[t]
\includegraphics[width=3.5in,keepaspectratio=true]{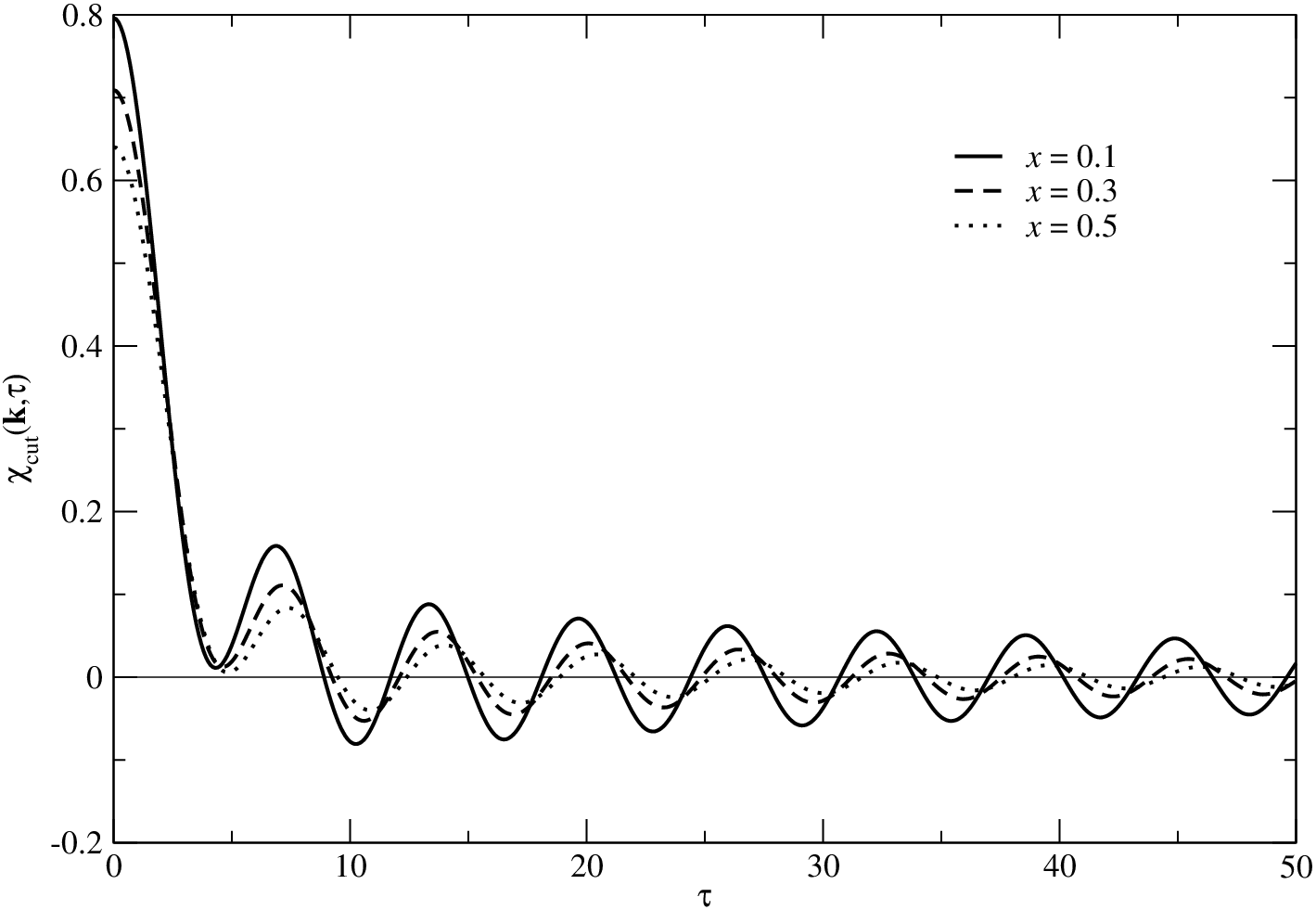}
\caption{$\chi_\mathrm{cut}(\bk,\tau)$ vs $\tau$ in the GL
region.}\label{fig:realtime}
\end{figure}

The Landau damping contribution to the real-time evolution of a
long-wavelength pion fluctuation $\chi_\mathrm{cut}(\bk,\tau)$ is
plotted in Fig.~\ref{fig:realtime}. The value at $\tau=0$ equals
$1-Z$ by the sum rule, thus it is clear from this figure that for
small $x$, namely $T \lesssim T_c$, the continuum contribution
dominates over that of the isolated pole. This figure also reveals
several time scales: a rapid damping as a function of the scaled
variable $\tau = kt$ towards an oscillating amplitude, which is of
the same order as the isolated pole contribution $Z\ll 1$ and
oscillates with a frequency $\alpha \simeq 1$ and slowly damped
out on longer time scale . The frequency of these oscillations is
clearly determined by the sharp rise of the Landau damping
contribution near $\alpha =1$.

\vspace{2mm}

Although for $T<T_c$ the pion is a stable quasiparticle, its
oscillating amplitude $Z$ vanishes as the critical temperature is
approached thus near the critical point the initial amplitude is
completely damped out by Landau damping. As $T\to T_c$ from below
the amplitude of long-wavelength pion fluctuations is damped by
Landau damping on a scale $\tau_*$ which is numerically
insensitive to $x$ for $x \lesssim 0.2$ and is approximately
$\tau_* \sim 5$ for the values of coupling and cutoff given by
eq.~(\ref{parameters}). In terms of the physical time, we see that
the amplitude of long-wavelength pion fluctuations relaxes on a
time scale $t_\mathrm{rel}(k)= \tau_*/k$, where $\tau_*\sim 5 $ is
numerically found to be rather insensitive to $x$. Thus relaxation
of long-wavelength pion fluctuations near the critical point are
\emph{critically slow}.

\subsubsection{Dynamics at the critical point}

As remarked above, the self-energies feature a logarithmic
threshold divergence at the critical point. Indeed, for $x\to 0$
the real and imaginary parts of $H(\alpha+i\epsilon)$ can be
approximated, respectively, as
\begin{eqnarray}
H_\mathrm{R}(\alpha,x)&\stackrel{x\to 0}{\simeq}&
\frac{1}{4\pi^2}\bigg[\int_0^{\bar{\Lambda}}
\frac{dz}{z}\tanh\frac{z}{2}- \bigg(1-\frac{\alpha}{2}
\ln\left|\frac{\alpha+1}{\alpha-1}\right|\bigg)
\tanh\frac{\bar{\Lambda}}{2}\bigg],\nn\\
H_\mathrm{I}(\alpha,x)&\stackrel{x\to 0}{\simeq}
&\frac{\alpha}{8\pi}\,\Theta(1-\alpha^2)\tanh\frac{\bar{\Lambda}}{2}.
\end{eqnarray}
Whereas the analytically continued inverse pion propagator
vanishes as $\alpha^2\to 1$ (with a logarithmic divergent slope
for the real part), the values $\alpha= \pm 1$ ($\omega = \pm k$)
are longer identified with quasiparticle poles but the endpoints
of the logarithmic branch cut. This threshold singularity is
reminiscent to those in QED and in equilibrium critical
phenomena\cite{boyan}. In the former case the self-energy of a
charged particle has a  logarithmic infrared divergence near the
threshold arising from the emission of soft gauge bosons. In the
latter case logarithmic infrared divergences appear in correlation
functions near the critical point at the upper critical dimension.
The emergence of infrared divergences in both cases indicates the
breakdown of perturbation theory and calls for a nonperturbative
resummation of the infrared divergences\cite{boyan}.

To display the logarithmic threshold divergence in an explicit
manner, let us focus on the Laplace transform of the inverse pion
propagator [see (\ref{Dinverse})], which near the threshold at the
critical point can be approximated as
\begin{eqnarray}\label{logs}
D_\pi^{-1}(k,s)&=& g \,\mathcal{C}\,\bar{k}^2
(1+\xi^2)\left[1+i\xi\lambda\ln\frac{i\xi+1}{i\xi-1}\right]\nn\\
&\stackrel{\xi^2\to -1}{\simeq}&g \,\mathcal{C}\,\bar{k}^2
(1+\xi^2)\left\{1-\lambda\ln[-(1+\xi^2)]+\cdots\right\},
\end{eqnarray}
where
\begin{equation}
\mathcal{C}=H_\mathrm{R}(0),\quad
\lambda=\frac{1}{8\pi^2\mathcal{C}}\tanh\frac{\bar{\Lambda}}{2}\approx
0.683,
\end{equation}
and the dots denotes terms that are formally of order $\lambda$
but remain finite in the limit $\xi^2\to -1$. Such logarithmic
threshold divergence at the critical point is akin to that found
in Ref.~\cite{boyan}. In this reference a finite-temperature
version of the renormalization group was introduced to provide a
resummation of these threshold divergences which leads to an
exponentiation and anomalous dimensions near threshold.

The logarithmic threshold divergence featured by the pion
propagator (\ref{logs}) \emph{suggests} the buildup of an
anomalous dimension. The results of Ref.~\cite{boyan} would
suggest that the leading logarithmic enhancement of the pion
self-energy at threshold eq.~(\ref{logs}) could exponentiate
leading to a resummed inverse pion propagator near the pion mass
shell
\begin{equation}
D_\pi^{-1}(k,s)\simeq - g\,\mathcal{C}\,\bar{k}^2
\left[-(1+\xi^2)\right]^{1-\lambda}\quad\mbox{for}\quad \xi^2\to
-1,
\end{equation}
which, after analytic continuation $i\xi\to \alpha+i\epsilon$,
gives
\begin{eqnarray}
D_\pi^{-1}(k,\omega)&\simeq& -g\,\mathcal{C}\,\bar{k}^2\,
|1-\alpha^2|^{1-\lambda}\big\{\cos[\lambda\pi\Theta(1-\alpha^2)]\nn\\
&&+i\,\mbox{sgn}(\alpha)\,\sin\lambda\pi\,\Theta(1-\alpha^2)\big\}
\quad\mbox{for}\quad \alpha^2\to 1.
\end{eqnarray}
Such an exponentiation \emph{if correct} would reveal a new
dynamical exponent (anomalous dimension) $\lambda$, which however
is numerically large to be justified by a perturbative approach.

Obviously at this stage we can only \emph{speculate} that such an
exponentiation actually emerges based on the study of
Ref.~\cite{boyan}. A full analysis of whether a finite-temperature
renormalization group approach can be systematically implemented
in this theory, remains  at this stage an open question that
deserves further study and which is clearly beyond the scope of
this article.

\subsubsection{Non-locality of the real-time effective action}

An important aspect of the pion self-energy is that it is
non-analytic for fixed $k$ as $\omega \rightarrow 0$. In
particular this non-analyticity of the function $H(\alpha)$
entails that the limits $k\rightarrow 0$ and $\omega \rightarrow
0$ \emph{do not commute}. Furthermore, this branch cut prevents a
well-defined expansion in powers of $k$ or $\omega$, which in turn
translates in that the effective action is \emph{non-local} in
space and time  and cannot be expressed as a \emph{gradient
expansion} in terms of space and time derivatives of the fields.
This non-locality is similar to that in the hard thermal loop
effective action in gauge theories~\cite{brapis,lebellac} which
also originates in Landau damping.

To highlight this point more clearly, we focus on the limits
$k,\,\omega \rightarrow 0$ of the retarded inverse pion propagator
[see eqs.~\eqref{Dinverse}]
\begin{equation}
D^{-1}_\mathrm{ret}(k,\omega)=
\widetilde{D}^{-1}_\pi(k,s=i\omega+\epsilon)=
-(\omega^2-k^2)\left[H_\mathrm{R}(\alpha,x)+iH_\mathrm{I}(\alpha,x)\right].
\end{equation}
We find the following limits
\begin{equation}\label{lims}
D^{-1}_\mathrm{ret}(k,0) = k^2H_\mathrm{R}(0),\quad
D^{-1}_\mathrm{ret}(0,\omega)= -\omega^2 H_\mathrm{R}(\infty),
\end{equation}
with $H_\mathrm{R}(0)$ and $H_\mathrm{R}(\infty)$ slowly varying
functions of the temperature near $T_c$. Their values at $T=T_c$
are given by
\begin{equation}
H_\mathrm{R}(0,T=T_c)= 0.015,\quad
H_\mathrm{R}(\infty,T=T_c)=0.04.\label{valuesTc}
\end{equation}
The limits (\ref{lims}) would seem to suggest that there are
stable long-wavelength pion fluctuations that propagate with a
dispersion relation
\begin{equation}
\omega^2 = v^2 k^2,\quad v^2=
\frac{H_\mathrm{R}(0)}{H_\mathrm{R}(\infty)}\approx 0.375.
\end{equation}
However, such conclusion is obviously incorrect, for $T<T_c$ there
are stable, long-wavelength pion fluctuations that propagate with
unit group velocity since the isolated pion poles correspond to
$\alpha^2=1$ [see eq.~(\ref{isopoles})]. Landau damping thus
prevents a local description of the low-energy effective action as
an expansion in spatiotemporal gradients. This point will be
important below when we discuss our results in light of previous
studies.

\subsection{Away from equilibrium: spinodal
instabilities}\label{sec:spino}

During a rapid phase transition, early stages of the
nonequilibrium dynamics is dominated by spinodal instabilities,
i.e., long-wavelength fluctuations are unstable and their
amplitude grows exponentially during the early
stages~\cite{spino}. These instabilities result in the emergence
of correlated domains, which in the case under consideration are
coherent domains of pions which have been identified as
disoriented chiral condensates~\cite{DCC}. Spinodal instabilities
emerge when the order parameter is not at the minimum of the free
energy for $T < T_c$ but is still ``rolling'' towards it.

In the case under consideration the equilibrium minimum in the GL
region is determined by eq.~(\ref{MFS}). Hence below the critical
temperature and in the GL region, we have
\begin{equation}
\frac{1}{\sigma_0}\frac{d\mathcal{V}_\mathrm{eff}(\sigma_0)}{d\sigma_0}
= b\left[\sigma^2_0 - \sigma^2_\mathrm{eq}(T)\right] \equiv
-\Delta^2(T),\label{unst}
\end{equation}
with $\Delta^2(T) >0$ in the \emph{spinodal region} $|\sigma_0| <
\sigma_\mathrm{eq}(T)$. The inverse pion propagator
(\ref{Dinverse}) in the GL region away from equilibrium can now be
written as
\begin{equation}
\widetilde{D}^{-1}_{\pi}(k,s)= g(\bar{s}^2+\bar{k}^2)
H(i\xi,x)-\Delta^2(T),\label{Dinverse2}
\end{equation}
where we have restored the term in \eqref{unst} which is absent in
equilibrium. It is clear from the inverse Laplace transform
expression that yields the real-time evolution (\ref{invlap}) that
\emph{real} poles in the Laplace variable $s$ correspond to
exponentially growing (or decaying) solutions. If the real part of
the pole is positive, the real-time evolution describes an
instability with a consequent exponential growth of small
amplitude fluctuations.

For spatially homogeneous pion fluctuations ($k= 0$) in the
spinodal region of the phase diagram with $|\sigma_0|<
\sigma_\mathrm{eq}(T)$ for which $\Delta^2(T)>0$, the inverse pion
propagator becomes
\begin{equation}
\widetilde{D}^{-1}_{\pi}(0,s)= g\,s^2 I(T)-\Delta^2(T),
\label{Dinverse0}
\end{equation}
where
\begin{equation}
I(T)=\frac{1}{4\pi^2}\int_0^{\bar{\Lambda}}\frac{dz
z^2}{\epsilon^3}\tanh\frac{\epsilon}{2}~~;~~I(T_c)=
\frac{1}{4\pi^2}\,\ln
\left(\frac{2~\bar{\Lambda}~e^{\gamma}}{\pi}\right) .
\end{equation}
\noindent with $\gamma$ the Euler-Mascheroni constant. The pion
propagator has real poles at
\begin{equation}\label{sspino0}
s^2_*= \frac{\Delta^2(T)}{g\,I(T)},
\end{equation}
which describes one growing and one decaying solution. The growing
solution has growth rate given by (for $T < T_c$)
\begin{equation}
\Gamma_*(k=0,T) = \frac{\Delta(T)}{\sqrt{g\,I(T)}}\propto
\big[\sigma^2_\mathrm{eq}(T)-\sigma^2_0\big]^{\frac{1}{2}}.
\label{growthrate}
\end{equation}

The \emph{classical spinodal} corresponds to the region of order
parameter in the phase diagram for which the system is
thermodynamically unstable to long-wavelength
perturbations~\cite{spino}. This corresponds to the values of the
order parameter for which the zero momentum growth rate
$\Gamma_*(k=0,T)\neq 0$. Therefore we see that the \emph{classical
spinodal} line corresponds to $|\sigma_0| <
\sigma_\mathrm{eq}(T)$. This is the \emph{same} as the classical
spinodal in the mean field approximation for the GL effective
theory or the linear sigma model. This can be seen from the
Lagrangian density (\ref{NLSMF}). Since along the classical
spinodal $ (1/\sigma_0)~dV(\sigma_0)/d\sigma_0 < 0$, hence the
mass squared term for the pions is negative and small amplitude
pion fluctuations are unstable and grow
exponentially~\cite{spino}.

While obtaining the growth rate for arbitrary $k$ is an involved
numerical task  which is not very illuminating, we can learn much
by obtaining $\Gamma_*(k,T)$ in the long-wavelength limit
$k^2/\Delta^2(T)\ll 1$. This is achieved by expanding the function
$H(is/k)$ for \emph{large} values of the ratio $s/k$ and solving
for the position of the (real) pole in the variable $s$. This
expansion is valid since the function $H$ is analytic everywhere
away from the Landau damping cut along the imaginary axis in the
complex $s$-plane. After straightforward algebra we find
\begin{equation}\label{gammaofK}
\Gamma_*(k,T) =
\left[\frac{\Delta^2(T)}{g\,I(T)}-[1-J(T)]\,k^2+g\,
I(T)\,J(T)\frac{k^4}{\Delta^2(T)}
+\mathcal{O}\left(\frac{k^6}{\Delta^4(T)}\right)\right]^\frac{1}{2},
\end{equation}
where
\begin{equation}
J(T)= \frac{1}{24\pi^2}\int_0^{\bar{\Lambda}}\frac{dz
z^4}{\epsilon^4}\frac{1}{\cosh^2\frac{\epsilon}{2}}~~;~~J(T_c)=
\frac{1}{12\pi^2}.
\end{equation}

While the expression (\ref{gammaofK}) can be expanded in a power
series in $k^2$ in the long-wavelength limit, we have purposely
kept the powers of $k$ inside the square root to compare with the
growth rate predicted by the linear sigma model (\ref{NLSMF})
\begin{equation}\label{GammaLSM}
\Gamma_\mathrm{LSM}(k,T) = \left[ \left|\frac{1}{\sigma_0} \frac{d
V(\sigma_0)}{d\sigma_0} \right|-k^2 \right]^\frac{1}{2}.
\end{equation}
The corrections to the coefficient of $k^2$ as well as the higher
powers of $k$ in the growth rate (\ref{gammaofK}) all originate in
the non-local Landau damping contribution.

\section{Conclusions and summary of main results}\label{sec:conclu}

In this article we have studied the low-energy effective action
for meson fields in the two-flavor NJL constituent quark model.
The main goal and focus are to study  static and dynamical
phenomena near the critical point for the chiral phase transition
and explore whether the low-energy effective action describes the
same universality class as the O(4) linear sigma
model~\cite{wilraja} or the nonlinear sigma model~\cite{son}.

We have obtained the low-energy effective action by integrating
out the quark fields to lowest order in the loop expansion up to
quadratic order in the fluctuations around a mean field
configuration. In this constituent quark model, the meson degrees
of freedom emerge as scalar and pseudoscalar bound states of quark
and antiquarks and their space-time dynamics is completely
determined by the underlying quark model. Both the static as well
as the dynamical aspects where studied in the low-energy,
long-wavelength limit and in the GL region near the critical point
in the broken symmetry phase. The evaluation of the effective
potential (free energy for a mean field configuration) as well as
the self-energies for the scalar and pseudoscalar mesons allow us
to obtain the effective action from which we can extract the
static limit. The scalar and pseudoscalar self-energies are
obtained to lowest order in the loop expansion upon integrating
out the quark degrees of freedom. The imaginary parts of the
self-energies in the low-energy, long-wavelength limits are
completely determined by Landau damping resulting from quark
(antiquark) scattering in the medium. The integration of the quark
degrees of freedom in the \emph{real-time} effective action allows
us to obtain the equations of motion for scalar and pseudoscalar
(pion) meson fluctuations around the mean field and to study the
real-time evolution of small amplitude fluctuations as an initial
value problem. The study of the real-time dynamics near the
critical point allows us to explore \emph{dynamical critical
phenomena}.

Our conclusions and main results are the following:
\begin{itemize}
\item[(i)]{\textbf{Static critical phenomena}. We have established
that the low-energy effective action in the static limit up to
quadratic order in the fluctuations around the mean field, which
is obtained by integrating out the quark fields up to one loop,
coincides with the GL free energy including the \emph{spatial}
gradients in the long-wavelength limit. Thus the \emph{static
limit} of the NJL constituent quark model with two flavors is in
the same universality class as the O(4) linear sigma model
(Heisenberg ferromagnet) in agreement with the conjecture in
Ref.~\cite{wilraja}. }

\item[(ii)]{\textbf{Dynamical critical phenomena}. Dynamical
phenomena of pions near the critical point is studied by focusing
on the equations of motion for small amplitude pseudoscalar
fluctuations around a mean field configuration. The equations of
motion are solved by Laplace transform as an initial value
problem. For fluctuations around the equilibrium minimum of the
effective potential below the critical temperature, we find that
there are stable long-wavelength pion excitations corresponding to
poles away from the continuum in the pion propagators. The
dispersion relation of the stable pion excitations is
$\omega_{\pi}(k)=k$ and the residue at the pion pole vanishes as
$Z \propto 1/|\ln(1-T/T_c)|$ as $T\rightarrow T_c$ from below. The
amplitude of the pion fluctuations is Landau damped on a time
scale $t_\mathrm{rel}(k) = \tau_*/k$ with $\tau_*$ a slowly
varying function of temperature near $T_c$. Thus near the critical
temperature relaxation of long-wavelength pion fluctuations are
\emph{critically slow}. At the critical temperature, the isolated
pion pole merges with the Landau damping continuum resulting in a
logarithmic divergence on the pion mass shell similar to the
logarithmic enhancement found in Ref.~\cite{boyan}.  Based on the
renormalization group arguments of Ref.~\cite{boyan} we
\emph{conjecture} that such a logarithmic divergence on the pion
mass shell at the critical temperature is the harbinger of a
dynamical anomalous dimension $\lambda$, which however turns out
to be fairly large in this model, $\lambda\sim 0.683$.

We have studied the dynamics of fluctuations \emph{away} from the
equilibrium minima of the effective potential to assess the growth
rate of spinodal fluctuations. We find that the \emph{classical
spinodal} line coincides with that of the mean field description
of the GL linear sigma model. However, the growth rate for
long-wavelength spinodal fluctuations is different from those
obtained from the GL effective theory. The Landau damping
contribution to the self-energy leads to a richer wavelength
dependence of the growth rate of small amplitude spinodal
fluctuations.}
\end{itemize}

Thus we conclude that whereas \emph{static} critical phenomena is
described by the universality class of the O(4) Heisenberg
ferromagnet (linear sigma model) near the critical temperature,
dynamical critical phenomena is \emph{not} in the same
universality class. We also conclude that below the critical
temperature there are stable pion excitations with a dispersion
relation $\omega_{\pi}(k)=k$, namely with unit group velocity.
This result is in disagreement with those found in
Ref.~\cite{son}. We argue that the source of disagreement is
likely to be found in the derivative expansion of the effective
action as proposed in  Ref.~\cite{son}. We argue that such
derivative expansion leading to a local effective action is not
reliable because the Landau damping contribution to the pion
self-energy introduces a cut in the complex frequency plane along
the real axis, which at the critical point is in the region $-k <
\omega < k$. This cut results in that the real-time effective
action is \emph{non-local} and cannot be expanded in
spatio-temporal derivatives. The results of our one loop
calculation must be \emph{compared with the results of the
one-loop calculation} in the linear sigma model in the first
article of ref.\cite{son} where the limits $\vec{k}\rightarrow 0,
\omega=0$ and the limit $\omega \rightarrow 0, \vec{k}=\vec{0}$
are studied. The Landau damping cut for $-k \leq \omega \leq k$
where the pion propagator is non-analytic makes these two limits
not commutative. We suspect that this non-analyticity is at the
heart of the discrepancy between our results and those of
ref.\cite{son}, at least at the one-loop level which is common to
both studies.

Such a non-analytic structure below the light cone is akin to that
found in the hard thermal loop limit of gauge theories which also
arises precisely from Landau damping~\cite{brapis,lebellac}. At
the critical point we find a logarithmic divergence on the pion
mass shell, again originating in the Landau damping processes.
These logarithmic divergences are akin to those found in
Ref.~\cite{boyan} and suggest the buildup of  (in this case a
large) dynamical anomalous dimension.

\textbf{Caveats: beyond lowest order}. Our results are based on
integrating out the quark degrees of freedom at lowest order in
the loop expansion. Given that the effective coupling is not small
the validity of such an expansion can be called into question.
Such expansion becomes exact in the large-$N_f N_c $ limit,
however, taking $N_f N_c$ to be large for fixed coupling $G^2$ one
can see from eq.~(\ref{Tcfin}) that $T_c \approx \Lambda$ and the
regime of validity of the constituent quark model as a description
of the critical properties must be re-assessed. In the linear (or
nonlinear) sigma model effective field theory, the self-energy
corrections arise from interaction terms between scalar and
pseudoscalar fluctuations. For example some of the results of
Ref.~\cite{son} are based on a one-loop contribution to the pion
self-energy from the $\delta$-$\pi$-$\pi$ vertex below $T_c$. One
could obtain the effective $\delta$-$\pi$-$\pi$ vertex from the
constituent quark model also in one-loop approximation by going
beyond quadratic order in the fluctuations. Such vertex
necessarily must be proportional to $m$ by chiral symmetry. A
local limit of such vertex can be obtained by taking first the
static and then the long-wavelength limit. However, using such
local vertex to calculate real-time dynamics would not be
consistent and the non-local form factor of the vertex must be
taken into account to calculate the corrections to the pion
self-energies. Such calculation is clearly much more complicated
than that in the local effective field theory because of the
momentum and frequency dependence of the vertex are likely to
change the results of Refs.~\cite{son,boyan}.

At the order considered, below the critical temperature the pion
pole is isolated, this is a consequence that to this leading order
there are no scattering contributions and therefore the pion does
not acquire a collisional width. Of course this situation will
change beyond this order, in particular corrections to next to
leading order in the large $N_c N_f$ limit will lead to scattering
corrections and the pion will acquire a collisional width, whose
behavior near the critical point must be studied. The collisional
width of the pion is an important ingredient in the manifestation
of the axial anomaly at finite temperature\cite{anomaly} whose
understanding thus requires to go beyond one-loop order, which is
the leading order in the large $N_c N_f$ limit. Furthermore, as we
mentioned in the introduction, a well known shortcoming of the NJL
model is its lack of confinement\cite{hatsuda,klevansky}, which is
manifest in that free quarks are present in the thermal bath below
the critical temperature for chiral symmetry. Clearly there are no
models, short of full QCD that can overcome this problem.


The logarithmic threshold enhancement near the critical point
suggests that a resummation scheme such as a renormalization group
 program as proposed in ref.\cite{boyan} must be
implemented to approach the critical region in a reliable manner.

The study of the long-wavelength small frequency limit requires a
resummation of self-energy and vertex diagrams, just like in the
case of transport phenomena\cite{jeon,conductivity} as a
consequence of the emergence of ``pinch singularities''\cite{jeon}
which are manifest as secular terms in the perturbative expansion
in real time\cite{conductivity}. In ref.\cite{conductivity} it was
shown that such resummation can be implemented via a dynamical
renormalization group. It is clearly interesting and important to
study the long-wavelength and small frequency limit near the
critical point implementing these resummation schemes, perhaps
supplemented by a renormalization of the scattering amplitudes
near the critical point as suggested in ref.\cite{boyan}. Clearly
such program is well beyond the one loop approximation which
despite the fact of being the leading order in the large $N_c N_f$
limit, does not include self-energy and vertex corrections which
are necessary for a full resummation
program\cite{jeon,conductivity}.

The implementation of such program  is beyond the scope of this
study.

Thus  our study suggests that the \emph{static} critical phenomena
is well described by a GL low-energy effective theory, but that
the \emph{dynamical} aspects are richer, with  novel dynamical
critical phenomena whose universality must be studied in detail.
Our study at the one-loop level should  be taken as indicative of
novel dynamical phenomena emerging near the critical point and
calls for resummation program  perhaps by implementing a
renormalization group approach akin to that of Ref.~\cite{boyan}
but directly in the constituent quark model for a deeper
understanding of dynamical scaling and anomalous dimensions.

\begin{acknowledgements}
D.B.\ thanks the US NSF for support under grant PHY-0242134. The
work of S.-Y.W.\ was supported by the US DOE under contract
W-7405-ENG-36.
\end{acknowledgements}


\begin{thebibliography}{99}
\bibitem{pisarski}
R.D. Pisarski and F. Wilczek, Phys. Rev. D \textbf{29}, 338
(1984); F. Wilczek, Int. J. Mod. Phys. A \textbf{7}, 3911 (1992).

\bibitem{wilraja}
K. Rajagopal and  F. Wilczek, Nucl.Phys. \textbf{B399}, 395
(1993); Nucl. Phys. \textbf{B404}, 577 (1993);   K. Rajagopal, in
\emph{Quark-Gluon Plasma 2}, edited by R.C. Hwa (World Scientific,
Singapore, 1995).

\bibitem{karsch1}
For recent reviews, see F. Karsch, Nucl. Phys.  \textbf{A698}, 199
(2002); and references therein.

\bibitem{qgp}
H. Meyer-Ortmanns, Rev. Mod. Phys. \textbf{68}, 473 (1996); B.
M\"uller, \emph{The Physics of the Quark-Gluon Plasma}, Lecture
Notes in Physics Vol. 225 (Springer-Verlag, Berlin, 1985); L.P.
Csernai, \emph{Introduction to Relativistic Heavy Ion Collisions}
(John Wiley and Sons, Chichester, England, 1994); C.Y. Wong,
\emph{Introduction to High-Energy Heavy Ion Collisions} (World
Scientific, Singapore, 1994).

\bibitem{gellmann}
M. Gell-Mann and M. Levy, Nuovo Cimento \textbf{16}, 705 (1960).

\bibitem{GN}
D.J. Gross and A. Neveu, Phys. Rev. D \textbf{10}, 3235 (1974).

\bibitem{NJL}
Y. Nambu and G. Jona-Lasinio, Phys. Rev. \textbf{122}, 345 (1961);
\textbf{124}, 246 (1961).

\bibitem{hatsuda}
T. Hatsuda and T. Kunihiro, Phys. Rep. \textbf{247}, 241 (1994),
Phys. Rev. Lett. \textbf{55}, 158 (1985), Phys. Lett. B
\textbf{185}, 304 (1987), Phys. Lett. B \textbf{198}, 126 (1987);
S. Chiku and T. Hatsuda Phys. Rev. D \textbf{57}, 6 (1998); T.
Hatsuda, T. Kunihiro and H. Shimizu Phys. Rev. Lett. \textbf{82},
2840 (1999).

\bibitem{klevansky}
S. Klevansky, Rev. Mod. Phys. \textbf{64}, 649 (1992).

\bibitem{schwarz}
T.M. Schwarz, S.P. Klevansky, and G. Papp, Phys. Rev. C
\textbf{60}, 055205 (1999); O. Scavenius, A. Mocsy, I.N.
Mishustin, and D.H. Rischke, \textbf{64}, 045202 (2001).

\bibitem{chodos1}
A. Chodos, F. Cooper, W. Mao, H. Minakata, and A. Singh, Phys.
Rev. D \textbf{61}, 045011 (2000).

\bibitem{fodor} Z. Fodor and S. D. Katz, JHEP \textbf{203}, 14
(2002); Nucl.Phys.Proc.Suppl. \textbf{106 } 441 (2002); Heavy Ion
Phys. \textbf{18} 41 (2003); hep-lat/0402006 ; hep-lat/0401023 ;
Z. Fodor, Nucl.Phys. \textbf{A715}, 319 (2003); F. Csikor, G.I.
Egri, Z. Fodor, S.D. Katz, K.K. Szabo, A.I. Toth, hep-lat/0401022;
hep-lat/0401016, and references therein.

\bibitem{philipsen} E. Laermann and O. Philipsen, hep-ph/0303042;
O. Philipsen, hep-ph/0110051.

\bibitem{forcrand} Ph. de Forcrand and O. Philipsen  Nucl.Phys. \textbf{B673} 170
(2003); hep-lat/0309109; hep-ph/0301209.

\bibitem{karsch}  C.R. Allton, S. Ejiri, S.J. Hands, O. Kaczmarek, F. Karsch, E. Laermann, Ch. Schmidt, L. Scorzato; Phys.Rev. \textbf{D66}
074507 (2002).

\bibitem{DCC}
K. Rajagopal and F. Wilczek, Nucl. Phys. \textbf{B399}, 395
(1993); S. Gavin, A. Gocksch and R.D. Pisarski, Phys. Rev. Lett.
\textbf{72}, 2143 (1994);  D. Boyanovsky, H.J. de Vega, and R.
Holman, Phys. Rev. D \textbf{51} , 734 (1995); D. Boyanovsky, M.
D'Attanasio, H.J. de Vega, and R. Holman, \textbf{54}, 1748
(1996); F. Cooper, Y. Kluger, E. Mottola, and J.P. Paz, Phys. Rev.
D \textbf{51}, 2377 (1995);  F. Cooper, Y. Kluger, and E. Mottola,
Phys. Rev. C \textbf{54}  3298, (1996);  Y. Kluger, F. Cooper, E.
Mottola, J.P. Paz, and A. Kovner,  Nucl. Phys. \textbf{A590}, 581c
(1995);  J. Randrup, Phys. Rev. C \textbf{62}, 064905 (2000);
Nucl. Phys. \textbf{A616}, 531 (1997).

\bibitem{rajagopal}
K. Rajagopal, Nucl. Phys. \textbf{A680}, 211 (2000); B. Berdnikov
and K. Rajagopal, Phys. Rev. D \textbf{61}, 105017 (2000); M.
Stephanov, K. Rajagopal and  E. Shuryak, Phys. Rev. D \textbf{60},
114028 (1999);  J. Randrup, Phys. Rev. D \textbf{63}, 061901
(2001), Heavy Ion Phys. \textbf{9}, 289 (1999).

\bibitem{fotboy}
D. Boyanovsky, H.J. de Vega, R. Holman, and S.P. Kumar, Phys. Rev.
D \textbf{56}, 3929 (1997); Phys. Rev. D \textbf{56}, 5233 (1997).

\bibitem{chodos2}
A. Chodos, F. Cooper, W. Mao, and A. Singh, Phys. Rev. D
\textbf{63}, 096010 (2001); F. Cooper and V.M. Savage, Phys. Lett.
B \textbf{545}, 307 (2002).

\bibitem{koide}
T. Koide and M. Maruyama, nucl-th/0308025.

\bibitem{boyan}
D. Boyanovsky, H.J. de Vega, and M. Simionato, Phys. Rev. D
\textbf{63}, 045007 (2001); D. Boyanovsky, H.J. de Vega,
\textbf{65}, 085038 (2002);  D. Boyanovsky and H.J. de Vega, Ann.
Phys. (N.Y.) \textbf{307}, 335 (2003); D. Boyanovsky, H. J. de
Vega, R. Holman, M. Simionato , Phys.Rev. \textbf{D60}, 065003
(1999).

\bibitem{son}
D.T. Son and M.A. Stephanov, Phys. Rev. D \textbf{66}, 076011
(2002); Phys. Rev. Lett. \textbf{88}, 202302 (2002).

\bibitem{endnote}
In our definition of the NJL model given in eq.~(\ref{NJLlag}) the
four-Fermi coupling is $G^2/2$ instead of the definition in
Refs.~\cite{hatsuda,klevansky} and we consider massless quarks.

\bibitem{kapustabook}
J.I. Kapusta, \emph{Finite-Temperature Field Theory} (Cambridge
University Press, Cambridge, England, 1989).

\bibitem{spino}
D. Boyanovsky, D.-S. Lee, and A. Singh, Phys. Rev. D \textbf{48},
800 (1993); D. Boyanovsky, Phys. Rev. E \textbf{48}, 767 (1993);
D. Boyanovsky, H.J. de Vega, and R. Holman in \emph{Topological
Defects and the Non-Equilibrium Dynamics of Symmetry Breaking
Phase Transitions}, edited by Y.M. Bunkov and H. Godfrin (Nato
Science Series, Kluwer, 2000) [hep-ph/9903534]; and references
therein.

\bibitem{lebellac}
M. Le Bellac, \emph{Thermal Field Theory} (Cambridge University
Press, Cambridge, England, 1996).

\bibitem{ctp}
J. Schwinger, J. Math. Phys. \textbf{2}, 407 (1961); L.V. Keldysh,
Sov. Phys. JETP \textbf{20}, 1018 (1965); K.T. Mahanthappa, J.
Math. Phys. \textbf{47}, 1 (1963); \textbf{47}, 12 (1963); K.-C.
Chou, Z.-B. Su, B.-L. Hao, and L. Yu, Phys. Rep. \textbf{118}, 1
(1985).

\bibitem{tadpole}
See, for example, D. Boyanovsky, H.J. de Vega, R. Holman, and
D.-S. Lee, Phys. Rev. D \textbf{52}, 6805 (1995); D. Boyanovsky,
M. D'Attanasio, H.J. de Vega, and R. Holman, \textbf{54}, 1748
(1996); and references therein.

\bibitem{ivp}
D. Boyanovsky, H.J. de Vega, D.-S. Lee, Y.J. Ng, and S.-Y. Wang,
Phys. Rev. D \textbf{59}, 105001 (1999); S.-Y. Wang, D.
Boyanovsky, H.J. de Vega, D.-S. Lee, and Y.J. Ng, Phys. Rev. D
\textbf{61}, 065004 (2000); S.-Y. Wang, D. Boyanovsky, H.J. de
Vega, and D.-S. Lee, Phys. Rev. D \textbf{62}, 105026 (2000); and
references therein.

\bibitem{caldas}  H.C. de Godoy Caldas and  M. Hott,
Phys.Rev. \textbf{D67} 045011 (2003); H.C.G. Caldas, A.L. Mota,
M.C. Nemes, Phys.Rev.\textbf{ D63 } 056011 (2001);  H.C. de Godoy
Caldas, Nucl.Phys. \textbf{B623}  503 (2002).


\bibitem{brapis}
E. Braaten and R.D. Pisarski, Nucl. Phys. \textbf{B337}, 569
(1990);  \textbf{B339}, 310 (1990); R.D. Pisarski, Physica A
\textbf{158}, 146 (1989); Phys. Rev. Lett. \textbf{63}, 1129
(1989); Nucl. Phys. \textbf{A525}, 175 (1991).

\bibitem{anomaly} H.Itoyama and A. H. Mueller, Nucl. Phys.
\textbf{B218}, 349 (1983); S. P. Kumar, D. Boyanovsky, H. J. de
Vega, R. Holman, Phys. Rev. \textbf{D61} 065002 (2000).

\bibitem{jeon} S. Jeon, Phys. Rev. \textbf{D52}, 3591 (1995); S.
Jeon and L.G. Yaffe, Phys. Rev. \textbf{D53}, 5799 (1996).

\bibitem{conductivity} D. Boyanovsky, H. J. de Vega and S.-Y.
Wang, Phys. Rev. \textbf{D67}, 065022 (2003).



\end{thebibliography}
\end{document}